\documentclass[11pt]{article}
\usepackage{amsmath}
\usepackage{amssymb}
\usepackage{graphicx,psfrag,epsf}
\usepackage{enumerate,verbatim}
\usepackage{enumitem}
\usepackage{natbib}
\usepackage{mathalfa}
\usepackage{prodint}
\usepackage{multirow}
\usepackage{url} % not crucial - just used below for the URL 
\usepackage{bm}
\usepackage[utf8]{inputenc}
\usepackage{caption}
\usepackage[english]{babel}
\usepackage{amsthm}
\usepackage[colorlinks,citecolor=blue,urlcolor=blue]{hyperref}
\usepackage{longtable,booktabs}
\usepackage{mathtools}
\usepackage{color, colortbl}
\definecolor{Gray}{gray}{0.9}
\definecolor{Gray2}{gray}{0.7}
\newcolumntype{g}{>{\columncolor{Gray}}c}
\newcolumntype{G}{>{\columncolor{Gray2}}c}
\usepackage{rotating}
\usepackage{framed,lipsum}
\usepackage[flushleft]{threeparttable}
\usepackage{float}
\usepackage{xr}

\makeatletter
\newcommand*{\addFileDependency}[1]{% argument=file name and extension
  \typeout{(#1)}
  \@addtofilelist{#1}
  \IfFileExists{#1}{}{\typeout{No file #1.}}
}
\makeatother

\theoremstyle{definition}

\theoremstyle{remark}
\newtheorem{remark}{Remark}

\newtheorem{theorem}{Theorem}

\def\h2s{{\hspace{0.2in}}}
\newcommand{\beq}{\begin{equation}}
\newcommand{\eeq}{\end{equation}}
\newcommand{\beqn}{\begin{eqnarray}}
\newcommand{\eeqn}{\end{eqnarray}}

\newcommand{\domark}{%
  \vbox to 0pt{
    \kern-\dp\strutbox
    \smash{\llap{*\kern1em}}
    \vss
  }%
}
\usepackage[pagewise]{lineno}
\usepackage{framed,lipsum}

\usepackage{changebar} %%[xcolor]
\def\b0{\mathbf{0}}

\def\bbeta{\boldsymbol{\beta}}
\def\bsb{\boldsymbol{\beta}}

\def\bb{\mathbf{b}}
\def\bz{\mathbf{z}}

\def\cov{\mathrm{Cov}}

\def\bdelta{\boldsymbol{\delta}}

\def\bZ{\mathbf{Z}}

\def\pr{\mathrm{P}}

\def\bR{\mathbf{R}}
\def\sign{\mathrm{sign}}

\def\bSigma{\boldsymbol{\Sigma}}

\def\bS{\mathbf{S}}
\def\bs{\mathbf{s}}

\def\bu{\mathbf{u}}

\def\bU{\mathbf{U}}
\def\var{\mathrm{Var\,}}

\def\rT{\mathrm{T}}

\begin{document}
	\title{\bf  Inference for High Dimensional Censored Quantile Regression} % Via Fused Multi-sample Splitting}
	\author{Zhe Fei$^1$, Qi Zheng$^2$, Hyokyoung G. Hong$^3$,
		and Yi Li$^{4}$ \\
	    1. Department of Biostatistics, University of California, Los Angeles\\
	    2. Department of Bioinformatics and Biostatistics, University of Louisville \\
	    3. Department of Statistics and Probability, Michigan State University \\
	    4. Department of Biostatistics, University of Michigan}
	\date{}
	\maketitle

	\begin{abstract}
		
		With the availability of high dimensional genetic biomarkers, it is  of interest to identify heterogeneous effects of these predictors on patients'  survival, along with proper statistical inference. Censored quantile regression has emerged as a powerful tool for detecting heterogeneous effects of covariates on survival outcomes. To our knowledge, there is little work available to draw inference on the effects of high dimensional predictors for censored quantile regression. This paper proposes a novel procedure to draw  inference on all predictors within the framework of global censored quantile regression, which investigates  covariate-response associations over an interval of quantile levels, instead of a few discrete values. The proposed estimator combines a sequence of low dimensional model estimates that are based on multi-sample splittings and variable selection. We show that, under some regularity conditions, the estimator is consistent and  asymptotically follows a Gaussian process indexed by the quantile level. Simulation studies indicate that our procedure can properly quantify the uncertainty of the estimates in high dimensional settings. We apply our method to analyze the heterogeneous effects of SNPs residing in lung cancer pathways on patients' survival, using the Boston Lung Cancer Survival Cohort, a cancer epidemiology study on the molecular mechanism of lung cancer.

	\end{abstract}
	
	\noindent%
	\vfill
	\noindent%
	\textsc{Contact:} Yi Li, yili@umich.edu, Department of Biostatistics, University of Michigan, Ann Arbor, MI 48109.
	Supplementary materials for this article are available online. Please go to www.tandfonline.com/r/JASA.
	
	\vfill
	\noindent
	\textsc{Article History:} Received December 2019; Accepted July 2021.
	\def\spacingset#1{\renewcommand{\baselinestretch}%
		{#1}\small\normalsize} \spacingset{1}

	\spacingset{1.45} % DON'T change the spacing!

	\section{Introduction}

	Lung cancer presents much heterogeneity in etiology \citep{mckay2017large,dong2012potentially,huang2009genome}, and some genetic variants  may insert different impacts on different quantile levels of survival time.  
	For example, in the Boston Lung Cancer Survival Cohort (BLCSC) \citep{Christiani2017blcs}, a cancer epidemiology cohort of over 11,000 lung cancer cases enrolled in the Boston area since 1992, it was found that SNP {AX.37793583} (rs115952579), along with age, gender, cancer stage and smoking status, had heterogeneous effects on different quantiles of survival time. % (Figure \ref{snpplot}). \gh{I removed this figure here, otherwise the figure should be located here as well}
	  A total of $674$  patients in the study were genotyped, with the goal of identifying lung cancer survival-predictive SNPs.  
	Target gene approaches, which focus on SNPs residing in cancer-related gene pathways, are appealing for increased statistical power in detecting significant SNPs \citep{moon2003current,risch2008lung,ho2019machine},  and  the investigators have identified  SNPs residing in 14 well-known lung cancer-related genes \citep{zhu2017driver,korpanty2014biomarkers,yamamoto2008pik3ca,kelley2001genetic}.  
	{One goal was to investigate} whether and how each SNP might play a different role among the high-risk (i.e., lower quantiles of overall survival) and low-risk (i.e., higher quantiles of overall survival) cancer survivors.
	
	%Censored Quantile Regression (CQR) \citep[among others]{powell1986censored,portnoy2003censored,peng2008survival} has emerged as a useful alternative  strategy for survival analysis, such as the Cox model and accelerated failure time model. 
	%
	
	Quantile regression (QR) \citep{koenker1978regression} is a significant extension of classic linear regression.
		By permitting the effects of active variables to vary across quantile levels, quantile regression can naturally accommodate and examine the heterogeneous impacts of biomarkers on different segments of the response variable's conditional distribution.
		As survival data are subject to censoring and may be incomplete, QR methods developed for complete data may be unsuitable. Efforts have been devoted to developing censored quantile regression (CQR) \citep[among others]{powell1986censored,portnoy2003censored,peng2008survival},  which has become a useful alternative strategy to traditional survival models, such as the Cox model and the accelerated failure time model. 
QR has also been widely studied to accommodate high dimensional predictors. For example, \cite{wang2012quantile} dealt with variable selection using non-convex penalization; \cite{zheng2013adaptive} proposed an adaptive penalized quantile regression estimator that can select the true sparse model with high probability; and \cite{fan2014adaptive} studied the penalized quantile regression with a weighted $L_1$ penalty in an ultra-high dimensional setting. As to high dimensional CQR (HDCQR), \cite{he2013quantile} provided a model-free variable screening procedure for ultrahigh dimensional covariates, and \cite{zheng2018high} proposed a penalized HDCQR built upon a stochastic integral based estimating equation. However, most of the existing works in HDCQR were designed to select a subset of predictors and estimate the effects of the selected variables, instead of drawing inference on all predictors.

		Progress in high dimensional inferences has been made for linear and non-linear models  \citep{zhang2014confidence,buhlmann2014high,javanmard2014confidence,ning2017general,fei2019drawing,fei2021estimation}. 
{	For example, \cite{meinshausen2009p}  proposed to aggregate $p$-values from multi-sample splittings for high dimensional linear regression.
	Another line of works referred to as {\em post-selection inference} includes \cite{berk2013valid}, \cite{lee2016exact}, and \cite{belloni2019valid}, which provided  post-selection inference at fixed quantiles for complete data. However, these methods may not handle censored outcomes.}    
	 For censored median regression, \cite{shows2010sparse} provided sparse estimation and inference, but it cannot handle high dimensional data.
		
	We propose to draw inference on high dimensional  HDCQR based on a  splitting and fusing scheme, termed Fused-HDCQR. Utilizing a  variable selection procedure for HDCQR such as \cite{zheng2018high}, our method operates partial regression followed by smoothing. Specifically, partial regression allows us to estimate the effect of each predictor, regardless of whether or not it is chosen by variable selection. The fused estimator aggregates the estimates based on multiple data-splittings and variable selection, with a variance estimator derived by the functional delta method  \citep{efron2014estimation,wager2018estimation}.
{	To comprehensively assess the covariate effects on the survival distribution, we adopt a ``global'' quantile model \citep{zheng2015globally} with the quantile level varying over an interval, instead of a local CQR that focuses only on a few pre-specified quantile levels.  The global quantile model can address the molecular mechanism of lung cancer, our motivating disease,  that hypothesizes  that some genetic variants may cause heterogeneous  impacts on different but unspecified segments of survival distribution \citep{mckay2017large,dong2012potentially,huang2009genome}. } 

 % we do need to address the biological motivation here. also the the sentence has been paraphrased 

% no need to talk about implementation here as it is not easy to understand why global may have an easy implementation

%With respect to implementation, the existing local CQR methods require conditional kernel smoothing to handle a general conditionally independent censoring mechanism \citep{wang2009locally, leng2013quantile}. As kernel smoothing is likely to suffer from ``curse of dimensionality," it is nearly infeasible to use it for modeling high dimensional data. }

%  {Lung cancer, the disease that  motivates this project,   presents considerable heterogeneity in etiology \citep{mckay2017large,dong2012potentially,huang2009genome}}, and some genetic variants may insert different impacts on different but unspecified segments of survival time. Taken all together, we have  developed our method within the framework of global censored quantile regression models.} \gh{this is redundant as the very similar paragraph is used in the beginning of intro}

	Our work presents several advantages. First, compared to  high dimensional Cox models \citep{zhao2012principled,fang2017testing,kong2021high}, the employed HDCQR stems from the accelerated failure time model \citep{wei1992accelerated} and offers  straightforward {interpretations} \citep{hong2019quantile}.  Second, utilizing the global conditional quantile regression, it uses various segments of the conditional survival distribution to improve the robustness of variable selection and capture global sparsity. 
{Third, our splitting-and-averaging scheme avoids the technicalities of estimating the precision matrix by inverting a $p \times p$ Hessian matrix of the log likelihood, which is a major challenge for debiased-LASSO type methods {\citep{zhang2014confidence,van2014asymptotically}} and is even more so if we apply debiased-LASSO to the CQR setting.} 
	Finally, as opposed to post-selection inferences \citep[][among others]{belloni2019valid},  Fused-HDCQR  accounts for variations in model {selection} and draws inference for all of the predictors.
	
	The rest of the paper is organized as follows.
	Section \ref{ch3.2} introduces the method, and Section \ref{ch3.3} details the asymptotic properties. Section \ref{ch3.4} derives a non-parametric variance estimator, Section \ref{ch3.5} conducts simulation studies, and Section \ref{ch3.6} applies the proposed method to analyze the BLCSC data. The technical details, such as proofs and additional lemmas, are relegated to the online Supplementary Materials.
	
	\section{Model and Method}\label{ch3.2}
	
	\subsection{High dimensional censored quantile regression}
	
	Let $T$ and $C$ denote the survival outcome and censoring time, respectively. We assume that $C$ is independent of $T$ given $\widetilde{\bZ}$, a $(p-1)$-dimensional vector of covariates ($p>1$). Let $X=\min\{T,C\},\Delta = 1\{T\le C\}$, and $\bZ = (1,\widetilde{\bZ}^\rT)^\rT$, where $1\{\cdot\}$ is the binary indicator function. The observed data, $D^{(n)}=\{(X_i,\Delta_i,\bZ_i),i=1,\dots,n\}$, are $n$ identical and independently distributed (i.i.d.) copies of $(X,\Delta,\bZ)$.
	With $Y = \log T$, let $Q_Y(\tau|\bZ) = \inf\{t: \pr(Y\le t|\bZ)\ge \tau\}$ be the $\tau$-th conditional quantile of $Y$ given $\bZ$. A global censored quantile regression model stipulates
	\begin{equation}\label{cqr}
	Q_Y(\tau|\bZ) = \bZ^\rT\bbeta^*(\tau) ,\; \tau \in (0,1),
	\end{equation}
	where  $\bbeta^*(\tau)$ is a $p$-dimensional vector of coefficients at  $\tau$. %and $0<\tau_U<1$ is the upper bound for estimable quantiles subject to identifiability constraint caused by censoring. 
	%The theoretical constraints and the practical selection of $\tau_U$ is discussed in \cite{peng2008survival}. % $\bbeta^*(\tau)$ is assumed to be sparse, that is $q = \left| \bigcup_{\tau \in (0,\tau_U]}S_{\tau}\right| =o(n)$, where $S_{\tau} = \{j: \beta^*_j(\tau) \ne 0\}$ is the active set at $\tau$. Detailed assumptions on the sparsity will be explored in Section \ref{ch3.3}.
	We aim to draw inference on $\beta^*_j(\tau)$ for each $\tau \in (0,\tau_U]$ and for all $j\in \{1,\dots,p\}$, where
	$0<\tau_U<1$ is an upper bound for estimable quantiles subject to identifiability constraint caused by censoring \citep{peng2008survival}.
	
	Let $N(t) = 1\{\log X\le t,\Delta = 1\}$, $\Lambda_T(t|\bZ) = -\log (1 - \pr(\log T\le t |\bZ) )$, and $H(u) = -\log (1-u)$. Then, $M(t) = N(t) - \Lambda_T(t \wedge \log X|\bZ)$ is a martingale process under model \eqref{cqr} \citep{fleming2011counting}  and hence $E (M(t) |\bZ) = 0$. We use $N_i(t)$ and $M_i(t)$,  $i=1,\dots,n$, to denote the sample analogs of  $N(t)$ and $M(t)$.
	Let $\theta_{i} (\tau) = \bZ^\rT_{i}\bbeta(\tau)$ and 
	$$\bU_{n}(\bbeta,\tau) = n^{-1}\sum_{i=1}^{n}\bZ_i\left\{N_i\left(\theta_{i} (\tau)\right) - \int_{0}^{\tau}1\{\log X_i\ge \theta_{i} (u) \}dH(u) \right\}.$$ We denote by $\bu(\bbeta,\tau)$ the expectation of $\bU_{n}(\bbeta,\tau)$. 
	
	The martingale property implies $\bu(\bbeta^*,\tau) = 0$  with $\tau \in [0,\tau_U]$,  entailing an estimating equation with  $\tau\in(0,\tau_U]$:
	\begin{equation}\label{phee}
	n^{1/2}\bU_{n}(\bbeta,\tau) =n^{-1/2}\sum_{i=1}^{n}\bZ_i\left\{N_i\left(\theta_{i} (\tau)\right) - \int_{0}^{\tau}1\{\log X_i\ge \theta_{i} (u) \}dH(u) \right\}=0.
	\end{equation}
	The stochastic integral in (\ref{phee}) naturally suggests sequential estimation {with respect to} $\tau$. We define a grid of quantile values $\Gamma_m = \left\{\tau_0,\tau_1,\dots, \tau_m \right\}$ to cover the interval $[\nu,\tau_U]$, where $\tau_0 =\nu$ and $\tau_m = \tau_U$.  
	The assumption on the lower bound $\nu>0$  is made to circumvent the singularity problem with CQR at $\tau = 0$, as detailed in assumption \ref{a1}. In practice, $\nu$ is chosen such that only a small proportion of  observations are censored below the $\nu$-th quantile. %Therefore $\tau_L$ is motivated by scientific interests, while $\tau_0$ is the threshold by assumption. 
	
	Then,  $\acute{\bbeta}(\tau_k)$'s, the estimates of $\bbeta(\tau_k)$'s, $\tau_k \in \Gamma_m$, can be sequentially obtained by 
	solving 
	\begin{equation*}
	n^{-1/2}\sum_{i=1}^{n}\bZ_i\left(N_i\left(\theta_{i} (\tau_k) \right) - \sum_{r= 0}^{k-1} \int_{\tau_r}^{\tau_{r+1}}1\{\log X_i\ge \acute{\theta}_{i} (\tau_r)  \}dH(u) \right) = 0,
	\end{equation*}
	where $\acute{\theta}_{i} (\tau_k)= \bZ^\rT_{i}\acute{\bbeta}(\tau_k)$. 
	Due to the monotonicity of $\theta_{i}(\tau)$ in $\tau$, $\acute{\bbeta}(\tau_{k})$ can be solved efficiently via $L_1$-minimization.  And  $\acute{\bbeta}(\tau),~\tau\in [\nu, \tau_{U}]$, is defined as a right-continuous
	piece-wise constant function that only jumps at the grid points.
	It can be shown that  $\acute{\bbeta}(\tau)$ is uniformly consistent and converges weakly to a mean zero Gaussian process for $\tau \in [\nu,\tau_U]$ when $p=o(n)$.  More importantly, $\acute{\bbeta}(\tau)$ provides a comprehensive understanding of the covariate effects on the conditional survival distribution over the quantile interval $[\nu, \tau_{U}]$. We  incorporate  this sequential estimating procedure for low dimensional CQR estimation in our proposed method.
	
	In addition, our method  requires  dimension reduction, which 
	can be accomplished by  existing methods, including the screening method proposed by \cite{he2013quantile} and
	the penalized estimation and selection procedure developed by  \cite{zheng2018high}. Specifically, \cite{zheng2018high}   incorporated an  $L_1$ penalty into the stochastic integral based estimating equation in \eqref{phee} to obtain {an} L-HDCQR estimator, which achieves a uniform convergence rate of $\sqrt{q\log(p\vee n)/n}$, and results in ``sure screening'' variable selection with high probability, where $q$ is defined in condition \ref{a4}. \cite{zheng2018high} also  proposed
	an AL-HDCQR estimator by employing the Adaptive Lasso penalties, which  attains a uniform convergence rate of $\sqrt{q\log(n)/n}$ and  selection consistency.  
	%The estimation consistency of both L-HDCQR and AL-HDCQR heavily depend on the stringent choice of the  tuning parameters, which may be hard to optimize in practice.  On the other hand, the selection properties of {\bf those} two estimators are more achievable, {\bf which can be  useful to our proposed method shown in Section x.x}.
	
	\subsection{Fused-HDCQR estimator}
	
	Our proposed  Fused-HDCQR procedure  consists of multiple data splitting, selecting variables, fitting low dimensional CQRs with partitioned data, applying \emph{append-and-estimate} to all predictors, and aggregating those estimates.

	\begin{enumerate}
		\item  With the full data $D^{(n)}$, determine via  cross-validation the tuning parameter(s) $\lambda_n$ of $\mathcal{S}$, an HDCQR variable selection method.
		\item Let $B$ be a large positive integer. For each $b=1,\dots,B$, 
		\begin{enumerate}[label=(\roman*)]
			\item randomly split the data into equal halves, $D_{1}^{b}$ and $D_{2}^{b}$;
			\item on $D_{1}^{b}$, apply $\mathcal{S}$ with $\lambda_n$ on $[\nu,\tau_U]$, to select a subset of predictors, denoted by $\widehat{S}^b_{\lambda_n}$, or $\widehat{S}^b$ for short;
			\item on $D_{2}^{b}$, for each $j =1,\dots,p$, append  $j$ to $\widehat{S}^b$
			such that $\widehat{S}^b_{+j} = \{j\}\cup \widehat{S}^b$, 
			fit a partial CQR on the covariates indexed by $\widehat{S}^b_{+j}$, and denote their coefficient estimates by $\widetilde{\bbeta}_{\widehat{S}^b_{+j}}(\tau), \tau\in[\nu, \tau_{U}]$. Here, $\widetilde{\bbeta}_{\widehat{S}^b_{+j}}(\tau)$ is a
			right-continuous piecewise-constant function with jumps only at the grid points of $\tau_{k}\in \Gamma_{m}$;
			\item denote  by $\widetilde{\beta}_j^b(\tau) = \left(\widetilde{\bbeta}_{\widehat{S}^b_{+j}}(\tau) \right)_j$ the entry of $\widetilde{\bbeta}_{\widehat{S}^b_{+j}}(\tau)$ corresponding to $Z_j$.  
		\end{enumerate}
		\item Fusing: the final estimate of $\beta^*_j(\tau),\;\tau \in [\nu,\tau_U],\;  j=1,\dots,p$ is 
		\begin{gather}\label{sshdcqr}
		\widehat{\beta}_j(\tau) = \frac{1}{B} \sum_{b=1}^{B} \widetilde{\beta}_j^b(\tau).
		\end{gather}
	\end{enumerate}
	
	\begin{remark}
		We could select  different tuning parameters for $\mathcal{S}$ in each data split, but with much added computation. Our numerical evidence suggests that a globally chosen $\lambda_n$ work well.
	\end{remark}
	
	\begin{remark}
		Our procedure needs a variable selection procedure to reduce dimension. For example, L-HDCQR  selects a subset: $\{ j\in\{2,\cdots, p\}: \max_k |\hat{\gamma}_j(\tau_k)| > a_0, \tau_k \in \Gamma_m\},$ where $\hat{\gamma}_j(\tau_k)$'s are the L-HDCQR estimates,  $a_0>0$ is a predetermined threshold, and  $j$ starts with 2 as the intercept term (corresponding to $j=1$) is always included in the model. For the choice of variable selection methods, our experience suggests that we adopt the screening method in \cite{he2013quantile} for fast computation, use  L-HDCQR for detecting any non-zero effects in the quantile interval $[\nu,\tau_U]$, and choose AL-HDCQR if we opt to select fewer predictors.
	\end{remark}

{
	\begin{remark} \label{remark3}
	We select $\lambda_n$ by minimizing a $K$-fold cross-validation error defined by
	deviance residuals in the presence of censored outcomes  \citep{zheng2018high}.
	%With censored outcomes, we use the deviance residual to define the $K$-fold cross-validation error as in \cite{zheng2018high} and selected $\lambda_n$ by minimizing it. 
Specifically, we partition the data to $K$ folds, and let $\widehat{\bbeta}_{\lambda}^{(-k)}(\tau)$ be the penalized estimate of $\bbeta(\tau)$ using all of the data excluding the $k$-th fold with a tuning parameter $\lambda$ and $\tau\in [\nu,\tau_U]$, 
where $k = 1, \ldots, K$. Under the global CQR model \eqref{cqr}, we define the cross-validation error as
%\[\text{CV Error}(\lambda; \tau)=\sum_{k = 1}^{K} \sum_{i \in \text{fold } k  }\big|\hat{D}_{i}[\widehat{\bbeta}_{\lambda}^{(-k)}\{\tau\} ]\big|,
%\]
\begin{equation}
\label{eq:cv}
\text{CV Error}(\lambda)=\sum_{k = 1}^{K} \sum_{i \in \text{fold } k  }\int_{\nu}^{\tau_{U}}\big|{D}_{i}[\widehat{\bbeta}_{\lambda}^{(-k)}(\tau) ] \big| d\tau,
\end{equation}
where
$$
{D}_{i}[\bbeta(\tau) ]=\sign\big\{{M}_{i}(\bbeta(\tau) )\big\}\sqrt{-2{M}_{i}(\bbeta(\tau) )+\Delta_{i}\log\big\{\Delta_{i}-{M}_{i}(\bbeta(\tau) )\big\}}
$$
with ${M}_{i}(\bbeta(\tau))=N_{i}(\bZ_{i}^{T}\bbeta(\tau) )-\int_\nu^{\tau}1\{\log X_{i}\geq N_{i}(\bZ_{i}^{T}\bbeta(u) ) \}dH(u)-\nu$.
Here, $H(u)=-\log(1-u)$, $N_i(\cdot)$ is the counting process, and ${M}_{i}(\bbeta(\tau) )$ is the martingale residual under model \eqref{cqr} \citep{zheng2018high}.
\end{remark}
}

{
\begin{remark}
When carrying out quantile regression at each grid point, 
we  formulate it as a linear programming problem \citep{Koenker2005}, which  can be solved by a simplex algorithm with a  computational complexity of $O(n^2p)$ \citep{klee1972good}. Since our grid size is $O(n)$ and the number of resampling, $B$, is  $O(n)$, the computational complexity of our procedure is  $O(n^4 p)$.
\end{remark}
}
	
	\section{Theoretical Studies}\label{ch3.3}
	
	\subsection{Notation and regularity conditions}
	For any vector $\bdelta\in \bR^p$ and a subset $S\subset \{1,\dots,p\}$,  denote by $S^C$  its complementary set, and  define $\|\bdelta\|_{r,S} = \|\bdelta_S\|_{r}$, the $l_r$-norm of the sub-vector $\bdelta_S$, in which $\delta_{jS}=\delta_{j}$ if $j\in S$ and $\delta_{jS}=0$ if $j\in S^C$. We set the following conditions.
	
	\begin{enumerate}[label=(A\arabic*)]
		\item \label{a1} There exist a quantile $\nu$ and a constant $c>0$ such that $$ n^{-1}\sum_{i=1}^{n}1\left\{\log C_i\le \bZ^\rT_i\bbeta^*(\nu) \right\} (1-\Delta_i) \le cn^{-1/2}$$ holds for sufficiently large $n$.
		\item \label{a2} (\emph{Bounded observations}) $\|\bZ\|_\infty\le C_0$.  Without loss of generality,
		we assume $C_0 = 1$.  In addition, $E |\log X|< \infty$.
		\item \label{a3} (\emph{Bounded densities}) Let $F_T(t|\bZ) = \pr\left(\log T\le t |\bZ \right)$, $\Lambda_T(t|\bZ) = -\log \left(1 - F_T(t|\bZ) \right)$, $F(t|\bZ) = \pr\left(\log X\le t |\bZ \right)$, and $G(t|\bZ) = \pr\left(\log X\le t,\Delta = 1 |\bZ \right)$. Also, define $f(t|\bZ) = dF(t|\bZ)/dt$, and $g(t|\bZ) = dG(t|\bZ)/dt$. 
		\begin{enumerate}[label=(\alph*)]
			\item There exist constants $\underline{f}$, $\overline{f}$, $\underline{g}$ and $\overline{g}$ such that
			\begin{gather*}
			\underline{f}\le \inf\limits_{\bz,\tau \in [\nu,\tau_U]} f(\bz^\rT\bbeta^*(\tau)|\bz) \le \sup\limits_{\bz,\tau \in [\nu,\tau_U]} f(\bz^\rT\bbeta^*(\tau)|\bz)  \le \overline{f}, \\
			\underline{g}\le \inf\limits_{\bz,\tau \in [\nu,\tau_U]} g(\bz^\rT\bbeta^*(\tau)|\bz) \le \sup\limits_{\bz,\tau \in [\nu,\tau_U]} g(\bz^\rT\bbeta^*(\tau)|\bz)  \le \overline{g}.
			\end{gather*}
			\item There exist constants $\kappa>0$ and $A$ such that, when $ |t|\le \kappa$,
			\begin{gather*}
			\sup\limits_{\bz,\tau \in [\nu,\tau_U]} \left|f(\bz^\rT\bbeta^*(\tau)+t |\bz) -f(\bz^\rT\bbeta^*(\tau)|\bz) \right| \le A|t|, \\
			\sup\limits_{\bz,\tau \in [\nu,\tau_U]} \left|g(\bz^\rT\bbeta^*(\tau)+t |\bz) -g(\bz^\rT\bbeta^*(\tau)|\bz) \right| \le A|t|.
			\end{gather*}
		\end{enumerate}

		\item \label{a4} (\emph{Sparsity}) Assume $\log p =o(n^{1/2})$, and let
		$$S_{\tau} = \{j: \beta^*_j(\tau) \ne 0\},\quad S^* = \bigcup\limits_{\tau \in [\nu,\tau_U]} S_{\tau} = \left\{j: \sup\limits_{\tau \in [\nu,\tau_U]} |\beta^*_j(\tau)| > 0\right\}, \quad \text{and} \quad q = |S^*|.$$ %, and $S_* =  \bigcup\limits_{\tau \in [\nu,\tau_L)} \{j: \beta^*_j(\tau) \ne 0\}$. 
		%We assume $n/p = o(1)$, $\log p =o(n^{1/2})$, and $q = |S^*|$ satisfies $q \le K_1n^{c_1}$.
		Let $\widehat{S}$ be the index set of covariates selected by $\mathcal{S}$ with a tuning parameter $\lambda_n$.
		There exist constants $0\le c_1<1/3$, $c_2,\;K_1,\;K_2>0$ such that $q \le K_1n^{c_1}$,  $|\widehat{S}|\le K_1n^{c_1}$, and
		\begin{equation*}
		\pr\left( S^* \subseteq \widehat{S} \right)\ge 1 - K_2(p\vee n)^{-1-c_2}.
		\end{equation*}
		
		\item \label{a5} Let $\tilde{\mu}(\tau) = E \left[1\left\{\log X>\bZ^\rT\bbeta^*(\tau) \right\} \right]$. There exists a  constant $L>0$ such that $|\beta^*_j(\tau_1) - \beta^*_j(\tau_2) | \le L|\tau_1 - \tau_2|$ and $|\tilde{\mu}(\tau_1) - \tilde{\mu}(\tau_2) | \le L|\tau_1 - \tau_2|$, for all $\tau_1,\tau_2 \in (\nu,\tau_U]$ and $1\le j\le p$.
		\item \label{a6} (\emph{Bounded eigenvalues})  $\bdelta^{\rT}E[\bZ_i\bZ_i^\rT]\bdelta/\|\bdelta\|^2
		$ 
		is bounded below and above by $\lambda_{\min}$ and $\lambda_{\max}$, respectively, over
		$\|\bdelta\|_{0}\leq K_{1}n^{c_1},\bdelta\ne 0$, where $0<\lambda_{\min}<\lambda_{\max}$; \\
		(\emph{Nonlinear impact})
		$
		c_{2}:=\inf_{\|\bdelta\|_{0}\leq K_{1}n^{c_1},\bdelta\neq {\bf 0}}
		E[({\bf Z}_{i}^{\rT}\bdelta)^{2} ]^{3/2}/
		E[|{\bf Z}_{i}^{\rT}\bdelta|^{3} ]>0$.
		
		\item \label{a7}  $\Gamma_m$ is equally gridded with  $\tau_k - \tau_{k-1} =\epsilon_n= c_{0}n^{-1}$ for $\tau_k\in \Gamma_m$ ($k=1,\dots,m$) and 
		%The grid size satisfies    $\epsilon_{n}=c_{0}n^{-1} $
		 a constant $c_0>0$.
	\end{enumerate}
	
	{Assumption \ref{a1} requires  the number of censored observations below the $\nu$-th quantile not to exceed $cn^{1/2}$, which is satisfied  if the lower bound of  $C$'s support is greater than the lower bound of $T$'s support, a reasonable scenario in real applications.}
	As recommended in \cite{zheng2018high}, $\nu$ is chosen such that only a small proportion of the observed survival times below the  $\nu$-th quantile are censored.
	\ref{a2} assumes that the covariates are uniformly bounded. As pointed out by \cite{zheng2015globally},  the global linear quantile regression model is most meaningful when the covariates are confined to a compact set to avoid crossing of the quantile functions.
	\ref{a3} ensures the positiveness of $f(t|\bZ)$ between $\bZ^\rT\bbeta^*(\nu)$ and $\bZ^\rT\bbeta^*(\tau_U)$, which is essential for the identifiability of $\bbeta^*(\tau)$ for $\tau<\tau_U$. 
	\ref{a4} restricts the order of data dimensions, as well as the sparsity of $\bbeta^*(\tau)$, which is necessary for the convergence of the low dimensional estimator in (\ref{phee}) (Condition C4 in \cite{wang2012quantile}).
	\ref{a4} also characterizes the ``sure screening'' property by $\mathcal{S}$. 
	This asymptotic property  does not assess the  variability of  selection with a finite sample; it is  crucial to account for such variability for high dimensional inference \citep{fei2019drawing,fei2021estimation}. %%The sure screening property can be satisfied, for example,  by using the L-HDCQR procedure \citep{zheng2018high}. 
{Also, 
several variable selection methods for high dimensional CQR satisfy the sure screening property in \ref{a4} with additional mild  conditions.
\begin{itemize}
    \item L-HDCQR: by Corollary 4.1 of \cite{zheng2018high}, a \emph{beta-min} condition is required in addition to the set of conditions in this paper. Explicitly, there exist constants $C_1, C_2>0$, such that 
    $$\inf_{j\in S^*} \sup_{\tau \in [\tau_L, \tau_U] } |\beta^*_j(\tau)| > C_1\exp(C_2q\tau_U)\sqrt{q\log (p \vee n)/n} + L\sqrt{q}\epsilon_n.$$
    
    \item AL-HDCQR: by Corollary 4.2 of \cite{zheng2018high}, AL-HDCQR achieves the stronger \emph{selection consistency} property, which implies the sure screening property.
    
    \item Quantile-adaptive Screening: by Theorem 3.3 of \cite{he2013quantile}, with a proper threshold value in their technical conditions, the screening procedure achieves the sure screening property.
\end{itemize}
}

	\ref{a5} characterizes the smoothness of $\bbeta^*(\tau)$. %\ref{a7} requires the lower bound on the non-zero effects.
	{The eigenvalue condition in  \ref{a6}  is the sparse Riesz condition in \cite{zhang2008sparsity},  satisfied by many commonly used covariance structures, including the compound symmetry structure and the first order autoregressive structure (AR(1)) \citep{zhang2008sparsity}. Also, the nonlinear impact condition controls the  minoration of the quantile regression objective function by a quadratic function, as adopted in \cite{zheng2018high}, for establishing the consistency of L-HDCQR estimator.  The condition is  satisfied when the covariates $\bf Z_i$ have a log-concave density, which includes the commonly used normal distribution, Wishart distribution and Dirichlet distribution \citep{lovasz2007geometry}.}
%	\ref{a6} is analogous to the assumptions on the covariance structure in the  high dimensional literature \citep{zhao2006model,belloni2011l1,fan2014adaptive,van2014asymptotically}. As an extension to 	Condition C4 in \cite{peng2008survival}, it ensures the convergence of low dimensional CQR but with a  diverging number of covariates.
	\ref{a7} details the fineness of $\Gamma_m$, which renders an adequate  approximation to the stochastic
	integration in \eqref{phee}. 
	
	\subsection{Theoretical properties of Fused-HDCQR}
	%%{\bf the whole section seems to be focusing on how to theorem 1, which is only a lemma for the main theorem. Do we need to have such a detailed description?}
	
	We first extend the results in \cite{peng2008survival} from a fixed $p$ to a \emph{$p$-diverges-but-less-than-$n$} case. The results are novel and critical since we allow the true model size $q=|S^*|$ to increase with $n$, while the selected $\widehat{S}^b$'s in the fused procedure vary around $S^*$.
	%%Before moving to a high dimensional setting ($p>n$), we  show  the properties of coefficient estimators for the CQR with a diverging number of true predictors. 
	Specifically, we assume a subset $S\subset \{1,\dots,p\}$ in Theorems \ref{thm1} and \ref{lem1}, where $|S| \le K_1n^{c_1}$, $0\le c_1<1/3$ and $K_1 > 0$.
	%%The proofs are relegated to the Appendix. The lemmas consider a generic subset $S\subset \{1,\dots,p\}$ with $|S| \le K_1n^{c_1}$, $0\le c_1<1/2$ and $K_1\le 1$. %, and denote the partial data $D^{(S)} = \{(X_i,\Delta_i,\bZ_{S,i}),i=1,\dots,n\}$ that includes only covariates with indices $j\in S$. 
	Let $\acute{\bbeta}_S(\tau),\tau \in [\nu,\tau_U]$ be the  estimator from \cite{peng2008survival} of fitting the CQR with $\bZ_{S}$ over the $\tau$-grid $\Gamma_m$. 
	
	\begin{theorem}\label{thm1}
		{(Consistency with a diverging number of covariates)}
		Under Conditions \ref{a1} -- \ref{a7} and given a subset $S\subset\{1,\cdots, p\}$ such that $S^{*}\subseteq S$ and  $|S|\leq K_{1}n^{c_{1}}$,  there exist positive constants $\zeta_1$ and $\zeta_2$ such that 
		$$
		\sup_{\nu\leq \tau\leq \tau_{U}}\|\acute{\bsb}_{S}(\tau)-\bsb^*(\tau)\|
		\leq \zeta_{1}\exp(\zeta_{2}) (K_{1}n^{c_{1}-1}\log n)^{1/2}
		$$
		with probability at least $1-20c_{0}^{-2}K_{1}n^{c_{1}-2}$.
	\end{theorem}
	\begin{remark}
		From the proof of this theorem (in particular, the proofs
		of Propositions 1 and 2 in the Supplementary Materials that lead to this theorem), it can be seen that $\zeta_1$ and $\zeta_2$ do not depend on the choice of $S$ or $n$.  Thus, $\zeta_1$ and $\zeta_2$ are universal for all possible $S$ satisfying  $S^{*}\subseteq S$ and  $|S|\leq K_{1}n^{c_{1}}$.
	\end{remark}
	Next, we derive the weak convergence of $\acute{\beta}_{j}$ for any $j\in S$.
	\begin{theorem}  \label{lem1} 
		{(Weak convergence with a  diverging number of covariates)} Under Conditions \ref{a1} -- \ref{a7} and  given a  $S\subset\{1,\cdots, p\}$ such that $S^{*}\subseteq S$ and  $|S|\leq K_{1}n^{c_{1}}$, it holds that
		\begin{equation*}
		\sqrt{n}\left(\acute{\beta}_j(\tau) - \beta^*_j(\tau) \right)
		\end{equation*}
		converges weakly to a mean zero Gaussian process for $\tau\in [\nu,\tau_U]$
		and any $j\in S$.
		
	\end{theorem}
	
	In high dimensional settings, the next theorem shows that the fused estimator  enjoys desirable theoretical properties.
	\begin{theorem}\label{thm2}
		Consider the Fused-HDCQR estimator in (\ref{sshdcqr}). Under assumptions \ref{a1} -- \ref{a7}, for any $j\in \{1,\dots,p\}$,
		\begin{equation*}
		\sqrt{n}\left(\widehat{\beta}_j(\tau) - \beta^*_j(\tau) \right)
		\end{equation*}
		converges weakly to a mean zero Gaussian process for $\tau\in [\nu,\tau_U]$.
		%	\end{enumerate}  
	\end{theorem}
	
{	Our framework  enables us to obtain the joint distribution of any $K$-dimensional estimated coefficients, where $K$ is a finite number.
Let $\mathcal{K}$ be the collection of the indices of $K$ covariates of interest. 
We can show that the weak convergence result of $\acute{\bbeta}_{\mathcal{K}}(\tau)$, a $K$-dimensional subvector of the oracle estimator, still holds for $\tau \in [\nu, \tau_U]$, that is, $\sqrt{n}(\acute{\bbeta}_{\mathcal{K}}(\tau)-\bbeta^{*}_{\mathcal{K}}(\tau)), \tau \in [\nu, \tau_U]$ converges to a $K$-dimensional Gaussian distribution at any $\tau \in [\nu, \tau_U]$. 
We only need to replace $\acute{\bbeta}_{j}(\tau)$ by  $\acute{\bbeta}_{\mathcal{K}}(\tau)$ in the proof of Theorem 2  in the Appendix and slightly modify the arguments accordingly. Consequently, the term I in the proof of Theorem 3 still converges weakly to a mean zero Gaussian distribution, while  the norms of items II and III are still $o_{p}(1)$. 
Therefore, Theorem \ref{thm2} still holds for any $K$-dimensional subvector of $\hat{\bbeta}_{\mathcal{K}}(\tau)$, i.e.,
$\sqrt{n}(\hat{\bbeta}_{\mathcal{K}}(\tau)-\bbeta^{*}_{\mathcal{K}}(\tau))
$ converges to a mean zero $K$-dimensional Gaussian distribution at any $\tau \in [\nu, \tau_U]$.
	}
	
	As shown in the proof,  the  covariance function of $\widehat{\beta}_j(\tau)$ depends on the unknown active set $S^*$, the unknown conditional density functions $f(t|\bZ)$ and $g(t|\bZ)$, and other unknown quantities. Thus,  it is not calculable.  The next section proposes an alternative  model-free variance estimator based on the functional delta method and the multi-sample splitting properties \citep{efron2014estimation,fei2021estimation}.

	\section{A Variance Estimator via the Functional Delta Method}\label{ch3.4}
	
	Let $J_{bi}\in\{0,1\}$ indicate whether the $i^{th}$ observation is in the $b^{th}$ sub-sample $D_{2}^b$, and $J_{\cdot i} =B^{-1} \sum_{b=1}^{B}J_{bi}$. For each $i=1,\dots,n$, we define the re-sampling covariance between $J_{bi}$ and $\widetilde{\beta}_j^b(\tau_k)$ at $\tau_k\in \Gamma_m$  as
	\begin{gather*}
	\bs_{ij}(\tau_k) = \frac{1}{B} \sum_{b=1}^{B}\left( J_{bi}- J_{\cdot i}\right) \left( \widetilde{\beta}_j^b(\tau_k) -\widehat{\beta}_j(\tau_k) \right). 
		\end{gather*}
	Define $\bS_{j}(\tau_k) = \left(\bs_{1j}(\tau_k),\, \bs_{2j}(\tau_k),\,\dots,\,\bs_{nj}(\tau_k) \right)^\rT$ and
	let $n_1 = |D_{2}^b|$. It follows that the  covariance  between $\widehat{\beta}_j(\tau_k)$ and $\widehat{\beta}_j(\tau_\ell)$ can be consistently estimated by 
	\begin{equation*} \label{ch3.cov}
	\widehat{\cov}_j(\tau_k,\tau_\ell)=\frac{n-1}{n} \left(\frac{n}{n-n_1}\right)^2 \sum_{i=1}^{n}\bs_{ij}(\tau_k)\bs_{ij}(\tau_\ell)
	= \frac{n(n-1)}{(n-n_1)^2} \bS_{j}^\rT(\tau_k) \bS_{j}(\tau_\ell),    %%\quad j\in [p], \tau_k \in \Gamma_m
	\end{equation*}
	where the multiplier $n(n-1)/(n-n_1)^2$ is a finite-sample correction for sub-sampling \citep{wager2018estimation}.
	In particular, by taking $\tau_\ell=\tau_k$, a variance estimator for $\widehat{\beta}_j(\tau_k)$ is 
	\begin{equation} \label{ch3.var1}
	\widehat{V}_j(\tau_k)= \frac{n(n-1)}{(n-n_1)^2}  \bS_{j}^\rT(\tau_k) \bS_{j}(\tau_k). 
	\end{equation}
	As in \cite{wager2018estimation}, it follows that $\widehat{V}_j(\tau_k)/\var\left(\widehat{\beta}_j(\tau_k) \right)\xrightarrow{p} 1$ with $n, B\rightarrow \infty$.
	Furthermore, for a finite $B$, we propose a bias corrected version of (\ref{ch3.var1}):
	\begin{gather} \label{ch3.var2}
	\widehat{V}_j^B(\tau_k) = \widehat{V}_j(\tau_k) - \frac{nn_1}{B(n-n_1)}\left\{B^{-1}\sum_{b=1}^{B}\left( \widetilde{\beta}_j^b(\tau_k) -\widehat{\beta}_j(\tau_k) \right)^2\right\},\quad \tau_k \in \Gamma_m.
	\end{gather}
	The correction term in \eqref{ch3.var2} is a suitable multiplier of the re-sampling variance of $\widetilde{\beta}_j^b(\tau_k)$'s, and converges to zero with $n\rightarrow \infty$ and $n_1=O(n)$. Thus, the two variance estimators in \eqref{ch3.var1} and \eqref{ch3.var2} are asymptotically equal. However, $\widehat{V}_j(\tau_k)$ in  \eqref{ch3.var1} requires  $B$ to be of order $n^{3/2}$ to reduce the Monte Carlo noise below the sampling noise, while $\widehat{V}_j^B(\tau_k)$ in \eqref{ch3.var2}  only requires  $B$ to be of order  $n$ to achieve the same  \citep{wager2014confidence}. 
	
	Since  $\widehat{\beta}_j(\tau) $ converges weakly to a Gaussian process by Theorem \ref{thm2}, and our variance estimators are consistent on the grid points, we define an asymptotic $100(1-\alpha)\%$ point-wise confidence interval for $\beta^*_j(\tau_k)$ at any $\tau_k \in \Gamma_m$ as
	\begin{equation*}
	\left( \widehat{\beta}_j(\tau_k) - \Phi^{-1}(1-\alpha/2) \sqrt{\widehat{V}_j^B(\tau_k)} \; , \; \widehat{\beta}_j(\tau_k) + \Phi^{-1}(1-\alpha/2) \sqrt{\widehat{V}_j^B(\tau_k)} \right),
	\end{equation*}
	where $\widehat{V}_j^B(\tau_k)$ is as defined in (\ref{ch3.var2}), and $\Phi$ is the standard normal cumulative distribution function. The $p$-value of testing $H_0: \beta^*_j(\tau_k) = 0$ for a $\tau_k \in \Gamma_m$ is
	\begin{gather*} 
	2\times \Big\{1-\Phi\left({|\widehat{\beta}_j(\tau_k)|/ \sqrt{\widehat{V}_j^B(\tau_k)} }\right)\Big\}.
	\end{gather*}

%%%%%%%%%%%%%%%%%%%%%%%%%%%%%%%%%%%%%%%%%%%%%%%%%%%%%%%%%%%%%%%%%%%%%
	\section{Simulation Studies}\label{ch3.5}
	
{
	{In various settings, we compare the proposed method, Fused-HDCQR (referred to as ``Fused'' in the tables and figures hereafter),  with some competing methods  in quantile regression or high dimensional inference. These methods include   
 \cite{wang2012quantile} (``W12'') and  \cite{fan2014adaptive} (``F14'') for quantile regression; 
	    \cite{zheng2018high} (``Z18'') for censored quantile regression;  and  \cite{meinshausen2009p} (``M09'') for inference with aggregated $p$-values from multi-sample splittings.	 

In the simulations and the later data analysis, we choose L-HDCQR  described in Section~\ref{ch3.3} as the variable selection tool for Fused-HDCQR. We  also explore the feasibility of using other  alternatives for variable selection, such as \cite{fan2009ultrahigh} (``F09'') and M09. }}

	    %\item Variable selection performance between different data-splitting schemes: 
	    	%\begin{itemize}
	    %\item \cite{fan2009ultrahigh} (F09) %which takes the intersection of variables selected by SIS on the data from multiple  data split 
	    %\item  \cite{meinshausen2009p} (M09) % which  selects variables based on the aggregated $p$-values from the mult-sample splittings \citep{meinshausen2009p} 
	    %\item   \cite{zheng2018high}  (L-HDCQR) %which is used as a screening tool in the first step of Fused-HDCQR.
		%\end{itemize}
	%\end{itemize}

	When implementing  Fused-HDCQR, we 
specify the number of splits as $B=300$, the quantile interval  as $[\nu,\tau_U] = [0.1, 0.8]$, and the grid length as  $m = n/\log p$.
{
As regards the selection of tuning parameters, Theorems \ref{thm1} and  \ref{lem1} suggest that our procedure  not be sensitive to tuning parameters as long as  they can ensure sure screening. In practical settings, we recommend to select tuning parameters by minimizing the  5-fold cross-validation error as in \eqref{eq:cv}, which may help achieve sure screening and works well in our simulations. }
	%%We first consider an example with true $\bbeta^*(\tau)$ invariant in $\tau$, then explore the examples with   $\beta^*_j(\tau)$'s changing with $\tau$.% Our last example concerns the comparison of the two methods.
	We study the following  examples with sparse non-zero effects, some of which are heterogeneous.

	{\bf Example 1. } The event times are generated by
	\begin{equation*}
	\log T_i = \widetilde{\bZ}^\rT_i\bb + \varepsilon_i,\quad i=1,\dots,n,
	\end{equation*}
	where the coefficient vector $\bb$ is sparse with $b_{20} = {0.5}, b_{40} = {1}, b_{60} = {1.5}$, $b_j = {0}$ for all other $j$'s, and $\varepsilon_i \sim N(0,1)$. Therefore, the true coefficients satisfy $\bbeta^*(\tau) = (Q_\varepsilon(\tau),\bb^\rT)^\rT$ for all $\tau \in (0,1)$, where $Q_\varepsilon(\tau)$, the $\tau$-th quantile of the distribution of $\varepsilon$, is the intercept. The covariates $\widetilde{Z}_{j,i}$'s are  i.i.d. from $\mathrm{Unif}(-1,1)$ and are independent across $j\in \{1,\dots,p\}$. The censoring times are generated independently as $\log C_i \sim N(3,17.25)$,
	%+ N(-5,1) + N(8,0.25), 
	giving a censoring rate around $25\%$.  %%Two scenarios are examined, i) ii) $\widetilde{\bZ}_i$'s follow multivariate normal distribution $N_p(0,\bSigma)$, with  $\bSigma = (\sigma_{k\ell})_{p\times p},\;\sigma_{k\ell} = 0.5^{|k-\ell|}$ the AR(1) correlation structure. The censoring time is generated as $\log C_i = N(0,16) + N(-3,1) + N(7,0.25)$ if $\tilde{Z}_{1,i} \ge 0$, and $\log C_i = N(0,16) + N(0,1)$ if $\tilde{Z}_{1,i} < 0$, so that the censoring distribution is covariate dependent. The censoring rate is around $35\%$. 
	
	\noindent{\bf Example 2. }  The event times follow
	\begin{equation}\label{eg3}
	\log T_i = \widetilde{\bZ}^\rT_i\bb + 1.5\widetilde{Z}_{3,i}\varepsilon_i,
	\end{equation}	
	where $b_{20} = 1, b_{40} = 1.5, b_{60} = 2$, $b_j = 0$ for all other $j$'s, and $\varepsilon_i \sim N(0,1)$. We first generate $\acute{\bZ}_i \sim N_p(0,\bSigma)$ with an AR(1)  $\bSigma = (\sigma_{k\ell})_{p\times p}$, where $\sigma_{k\ell} = 0.3^{|k-\ell|}$, and then let $\widetilde{\bZ}_i = \acute{\bZ}_i$, except that the third covariate $\widetilde{Z}_{3,i} = |\acute{Z}_{3,i}| + 0.5$. Thus,  $\beta^*_1(\tau) = 0, \beta^*_4(\tau) = 1.5Q_\varepsilon(\tau)$, and $\beta^*_{j+1}(\tau) = b_j$, for all other $j$'s. The censoring times are generated independently as $\log C_i \sim  N(4,17.25)$, giving a censoring rate around $23\%$.
	
\begin{figure}[H]
			\centering
		\caption{Estimated heterogeneous effects and confidence intervals of Fused-HDCQR using Example 3: $\beta_2^*(\cdot)$ (left panel) and $\beta_{5}^*(\cdot)$ (right panel). From the top to the bottom are the plots for $(n,p)= (300,1000), (700, 1000)$ and $(700, 2000)$, respectively. \label{fig1}}
		\includegraphics[scale=0.6]{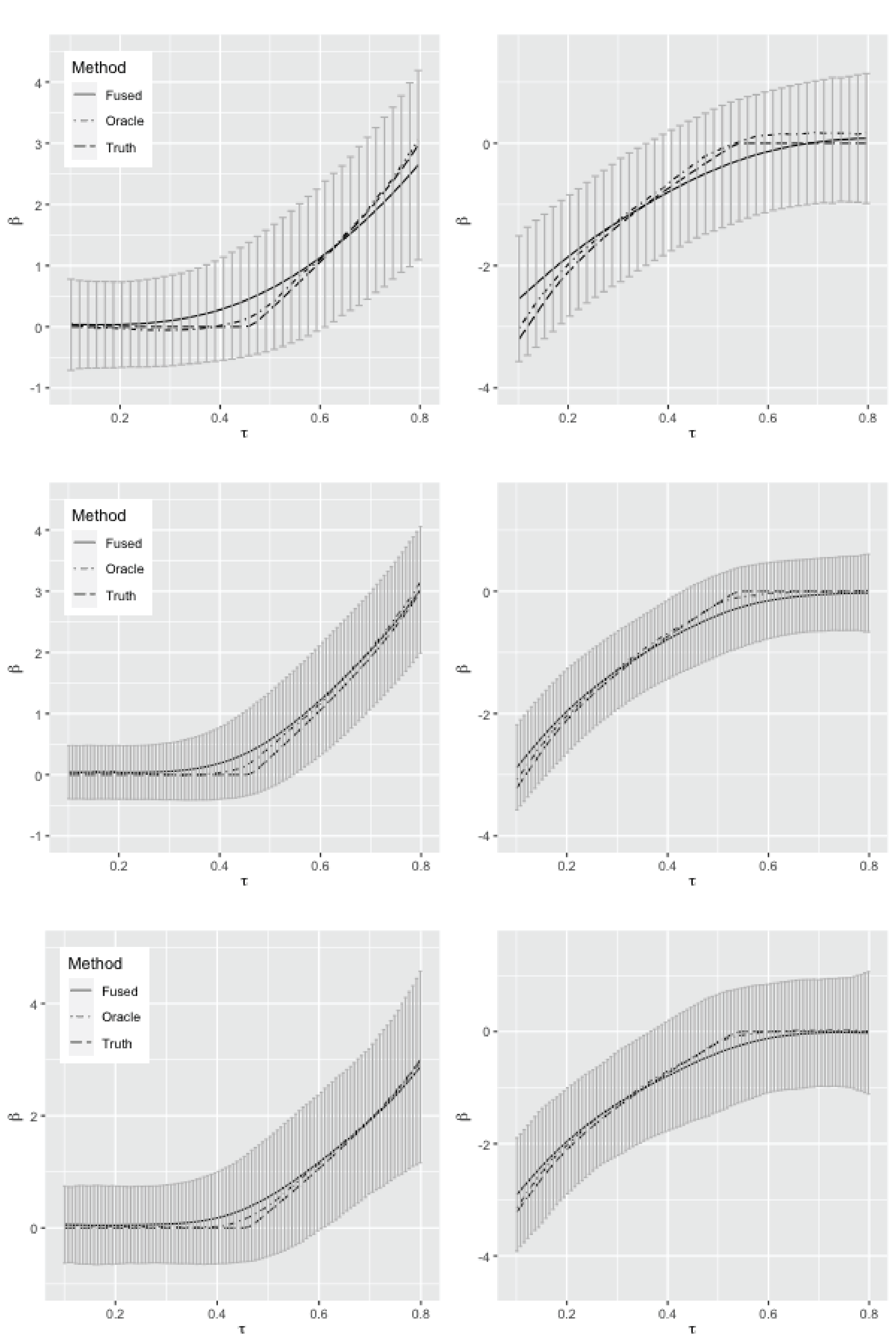}
	\end{figure}

	\noindent{\bf Example 3. }  The event times follow
	\begin{equation*}
	\log T_i = \widetilde{\bZ}^\rT_i\bb + \phi_1(\xi_i) \widetilde{Z}_{1,i} + \phi_{4}(\xi_i) \widetilde{Z}_{4,i},
	\end{equation*}	
	where $b_{20} = 1, b_{40} = 1.5, b_{60} = 2$, $b_j = 0$ for all other $j$'s, $\xi_i \sim N(0,1)$, and $\phi_1,\phi_{4}$ are monotone functions as the dashed lines in Figure \ref{fig1}, both are continuous with zero and non-zero pieces over $\tau$. We first generate $\acute{\bZ}_i \sim N_p(0,\bSigma)$ as in Example 2, and then let $\widetilde{\bZ}_i = \acute{\bZ}_i$, except that $\widetilde{Z}_{1,i} = |\acute{Z}_{1,i}| + 0.5$ and $\widetilde{Z}_{4,i} = |\acute{Z}_{4,i}| + 0.5$. Therefore, $\beta^*_1(\tau) = 0, \beta^*_2(\tau) = \phi_1(\tau), \beta^*_{5}(\tau) = \phi_{4}(\tau)$, and $\beta^*_{j+1}(\tau) = b_j$, for all other $j$'s. The censoring times are generated independently as $\log C_i \sim  N(6,17.25)$, which gives a censoring rate around $20\%$.

	 For each of these examples, we  set $(n,p)$=$(300,1000)$ and $(700,1000)$ to study the impacts of  the sample size and the number of variables on the performance,	and, in particular, how the methods fare when $p>n$.  In Example 3, which  mimics the real data example  in Section \ref{ch3.6} most closely, we have also explored $(n,p)=(700,2000)$, which is roughly equal  to the dimension of the real dataset.
For every parameter configuration, 	 a total of 100 independent datasets are generated, and we report the average  results based on these replications. {We choose 100 replications  because the penalized methods for high dimensional CQR  in general take much computing time   (Table \ref{simtab11}).}

%comparison between data-splitting method

%We first compare the results of variable selection between F09, the data-splitting based SIS in \cite{fan2009ultrahigh},  M09, the aggregated  $p$-value approach   in \cite{meinshausen2009p}, and Z18,  the  screening approach that we used for our proposed fused method in \cite{zheng2018high}, under Examples 1--3. 
{We first evaluate the feasibility of using  various variable selection tools for our proposed method.  Comparisons of true positives and false negatives among F09,  M09, and  L-HDCQR  under Examples 1--3 are reported in Table \ref{simtab1}.
	F09 presents a subpar performance because, by taking intersections of variables selected from different partitions of data, it tends to miss out some true signals and thus have fewer true positives. In contrast,   L-HDCQR  retains  more true positives 
than both F09 and M09, while having  more false positives.
   % Because our method requires to include the true signals with high probability in the selection step, even at the cost of some false positives, we have opted not to utilize F09 as the screening tool for our method.
        Because our method requires the variable selection step to include the true signals with high probability, even at the cost of some false positives, we use  L-HDCQR  as the screening tool for our method.}
    %In our simulations and data analysis, we have used Z18, which will also be referred to as L-HDCQR in this paper,  to screen variables because it achieves the sure screening property and has a better finite sample performance in containing the true signals.

%comparison between HDCQR

{
We next compare the performance of Fused-HDCQR with other high dimensional quantile regression methods at $\tau=.25, .5,. 75$ under  Example 1.
	%reports the bias of the oracle estimates (with the known  $S^*$), the proposed method and three penalized methods \citep{zheng2018high,fan2014adaptive,wang2012quantile}; the empirical standard deviation, estimated standard error and coverage probability of the proposed method; and the testing power of the proposed method and \cite{meinshausen2009p}. We have also compared the power of testing the signals between our method and Meinshausen et al. (2009). In addition, we have provided average standard errors (SE), empirical standard deviations (SD) and coverage probabilities of the confidence intervals derived with our method.
As a benchmark for comparisons,  we also compute the oracle estimates  based on the true model (with  $S^*$ known).
Since  W12, F14, and Z18 only provide coefficient estimates without standard errors (SEs), we only report the estimation biases for them, while reporting the average SEs, empirical standard deviations (SDs) and coverage probabilities of the confidence intervals  for our method. }
	Table~\ref{tab1} shows that Fused-HDCQR presents the smallest biases, which are comparable to those of the oracle estimates.  In contrast,  Z18  has smaller biases when the sample size is large, and  larger biases otherwise, while  W12 and F14 incur substantial biases since they are not designed for censored data.  Moreover,  the average SEs based on Fused-HDCQR agree with the empirical SDs of the estimates.  The consistent estimates of coefficients and SEs obtained by Fused-HDCQR lead to proper coverage probabilities around the $0.95$ nominal level.  {In addition, the coverage probabilities improve as $n$ increases. 
	
Table~\ref{tab1} also concerns the power for detection of signals. Since  W12, F14, and Z18 cannot draw inference
and, in general,  there is a lack of literature that deals with inference for HDCQR, we compare our method with the aggregated
$p$-value approach (M09) in the quantile setting, though M09 originated from linear regression. 
 The results indicate that Fused-HDCQR outperforms M09, presenting more power when the effect size is moderate or large. }

% Table \ref{tab2} summarizes the results from Example 2. Since \cite{wang2012quantile} and \cite{fan2014adaptive} are high dimensional quantile regression methods that do not account for censoring and have shown a subpar performance in Example 1, a comparison with them is omitted.	Similar observations can be made as in Example 1. Specifically for the heterogeneous effect $\beta^*_4(\tau) = 1.5Q_\varepsilon(\tau)$, our method  provides smaller biases than \cite{zheng2018high} and a higher power than \cite{meinshausen2009p}.
%	{
%	The reported selection frequency of each signal (over $B$ splits) suggests that most signals can be consistently selected as $n$ increases and the sure screening property in assumption \ref{a4} may be satisfied.
%	}

{Table \ref{tab2} summarizes the results from Example 2 with the heterogeneous effect $\beta_4$ varying with $\tau$.
We compare the estimation accuracy between Fused-HDCQR and  Z18, as well as  the statistical power between  Fused-HDCQR and M09.  Again, Fused-HDCQR  presents smaller biases than Z18 and a higher power than M09. 
{
{
To assess whether the tuning parameters selected as in Remark \ref{remark3} help the variable selection method (L-HDCQR), used by Fused-HDCQR, satisfy assumption \ref{a4} in Section~\ref{ch3.3}, we report the selection frequency of each signal variable in Table \ref{tab2} (and also in Table \ref{tab3}), and observe that  the selection frequency  increases as the sample size  increases,  hinting that  assumption \ref{a4} may be satisfied with these selected tuning parameters.
}
}
}
%	{
%	The reported selection frequency of each signal (over $B$ splits) suggests that most signals can be consistently selected as $n$ increases and the sure screening property in assumption \ref{a4} may be satisfied.
%	}	

{	
Table \ref{tab3} summarizes the results based on Example 3.  For the two heterogeneous effects $\beta_2$ and $\beta_{5}$ that vary with $\tau$,  their estimation biases of Fused-HDCQR become smaller and the estimated SEs are closer to the empirical ones as $n$ increases.  Figure \ref{fig1}  shows that the Fused-HDCQR estimates in general agree with the oracle estimates and the truth, except  at the non-smooth change points, and have narrower confidence intervals with a larger $n$, where the vertical bars are the average confidence intervals of the $\tau$ grid. }

%The biases are larger when $n=300$ and improve with increased $n$ and a finer grid for $\tau$. In addition, the true $\beta_2^*(\cdot)$ and $\beta_{5}^*(\cdot)$ are not smooth at their respective change points, which is around $\tau = 0.5$. Both the oracle and the proposed estimates are smooth everywhere and therefore less accurate at the change points. This may be improved with a even finer grid of $\tau$ but would increase the computation cost at the same time. Nevertheless, estimating heterogeneous effects with a non-smooth change point is more challenging than estimating fixed effects, but our confidence intervals still cover the truth with high probabilities. }

%%%%%%%%%%%%%%%%%%%%%%%%%%%%%%%%%%%%%%%%%%%

{In regards to the choice of $B$ in the variance computation, our numerical experience suggests that it may be sufficient to use a $B$ that is of the same order of the sample size, even when $n$ is less than $p$. This coincides with 
the note under \eqref{ch3.var2} that $B$ is
	only required to be of order $n$  to reduce the Monte Carlo noise below the sampling noise. 	 }

Finally, we compare the computation intensity among Z18, M09, W12, F14, and Fused-HDCQR under 
Example 1 and report in Table \ref{simtab11}  the computing time on average per dataset.
Our method is the most computationally intensive, because it involves multiple data-splittings and draws inferences on all of the $p$ coefficients. However, by  utilizing parallel computing, we have  managed to reduce the computational time to the same order of Z18, W12, and F14 that are based on penalized  regression. The R code used for generating the simulation results can be accessed via \url{https://github.com/feizhe/HDCQR_Paper}. 

	\section{Application to the Boston Lung Cancer Survival Cohort (BLCSC)}\label{ch3.6}
	
	Detection of molecular profiles related to  cancer survival  can aid personalized treatment in prolonging  patients' survival  and improving their quality of life.  In a subset of BLCSC samples, 674 lung cancer patients were measured with  survival times, along with  $40,000$ SNPs and clinical indicators, such as lung cancer subtypes (adenocarcinoma, squamous cell carcinoma, or others),  cancer stages (1-4),  age, gender, education level ( $\le$ high school or  $>$ high school) and smoking status (active or non-active smokers); see Table \ref{dem_blcs} for patients' characteristics.  The censoring rate was 23\% and a total of 518 deaths were observed during the followup period, with  the observed followup time varying from 13 to $8,584$ days. 
	
	We could have included all 40,000 SNPs in our analysis. However, for more statistical power, we opt for the targeted gene approach by focusing on 2,002 SNPs residing in  14 genes identified to be cancer related, namely, ALK,   BRAF,  BRCA1,   EGFR,  ERBB2,  ERCC1,   KRAS,    MET, PIK3CA,    RET,   ROS1,   RRM1,   TP53, and TYMS \citep{brose2002braf,toyooka2003tp53,paez2004egfr,soda2007identification}. Pinpointing the effects of individual loci within the targeted genes is helpful for understanding disease mechanisms \citep{evans2011interaction,d2019systematic}
	and designing gene therapies \citep{paques2007meganucleases,hanawa2004extended}.
	We also adjust for patients' clinical and environmental characteristics listed in Table~\ref{dem_blcs}, which gives  a total of $p=2,011$ predictors.   
	
	We apply Fused-HDCQR to compute the point estimates  (\ref{sshdcqr}) and the variance estimates (\ref{ch3.var2}).
	We set the quantile interval to be $[0.2,0.7]$, which is wide enough to  cover high- and low-risk groups and, in the meantime, ensures the quantile parameters be estimable in the presence of censoring  \citep{zheng2015globally}.  We choose the lower bound $\tau_0=\nu = 0.1$ to circumvent the singularity problem with CQR at $\tau = 0$, because few  ($<2\%$) observations are censored below the $0.1$-th quantile. 
	With $\epsilon_n = 0.01$, we form the $\tau$-grid $\Gamma_m$ of length $m = 61$. We set $B=750$ as the number of re-samples, which is  sufficiently large and comparable to the sample size.
	{To determine the tuning parameter $\lambda_n$ in L-HDCQR for selection, we use 5-fold cross-validation as specified in Remark \ref{remark3}.}
	
\begin{figure}
		\centering
		\caption{Estimated quantile-specific coefficients of the predictors in Table \ref{tab_blcs}.   \label{snpplot}}
		\includegraphics[scale=0.54]{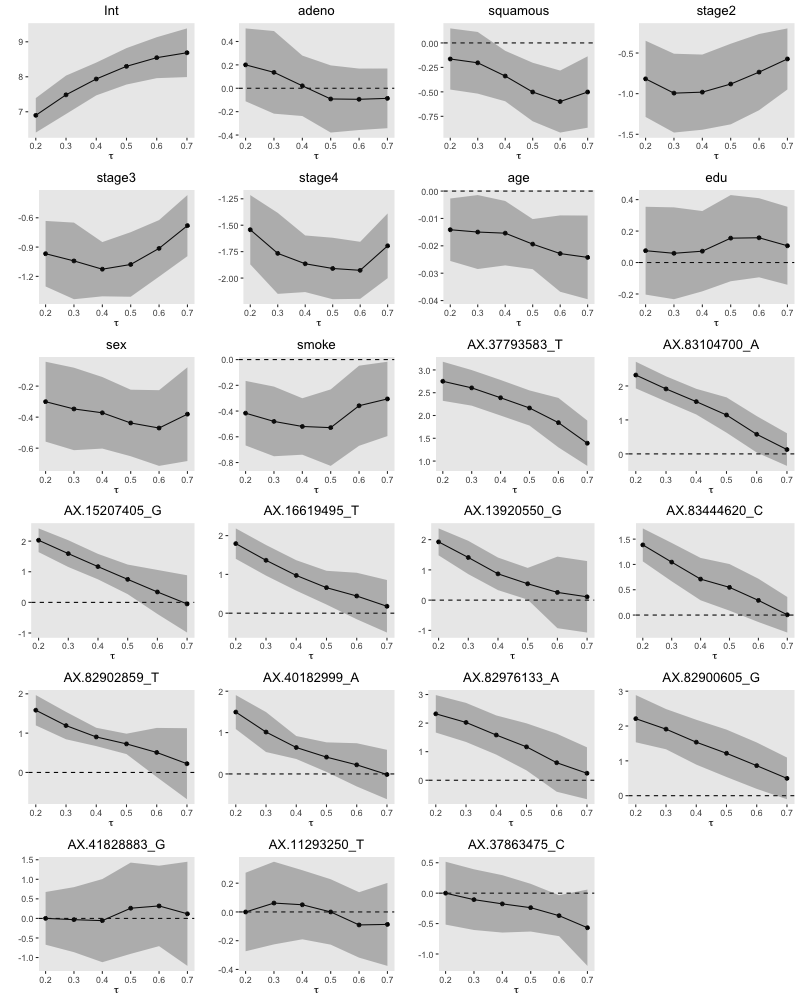}
	\end{figure}

	%, {\bf (need to discuss the robustness of the proposed Fused-HDCQR toward using different HDCQR approaches rather than Zheng et al. 2018). we applied the HDCQR method by \cite{zheng2018high} at  $\tau=\tau_0$ with 5-fold cross validation. However, we note that our results is robust toward the choice of the HDCQR method as shown in xxxxx.)}
	% Last, inferences were drawn based on the coefficient estimators and their standard errors. 
	For ease of presentation, we summarize the results evaluated at $6$  quantile levels, $\tau = 0.2,0.3,\dots,0.7$, instead of the whole grid $\Gamma_m$. To highlight the findings of the high-risk group,  we rank all SNPs based on their $p$-values at $\tau = 0.2$. In particular, after Bonferroni correction for multiple testing, there are $83$ significant SNPs for $\tau = 0.2$ with the overall type I error of $\alpha = 0.05$. 
	Our method estimates the coefficients and the $p$-values for \emph{all} predictors, and we only present the results for the patient characteristics, the top $10$ significant SNPs, and the $3$ least significant SNPs in Figure \ref{snpplot} and Table \ref{tab_blcs}. The estimated coefficient of active smoking drops from $-0.42$ ($p=0.0011$)  to $-0.53$ ($p=0.0005$) as $\tau$ changes from $0.2$ to $0.5$, and then increases to $-0.31$ ($p=0.038$) as $\tau$ changes to $0.7$, suggesting that active smoking might be more harmful to the high- and median-risk groups than the low-risk group of patients. %We suspected the reason in change of standard errors being that there were more data for middle quantiles.
	%Advanced cancer stages  had harmful effects on patients survival at medium quantiles. 
	The most significant SNP at $\tau=0.2$ is  AX.37793583\_T, which remains significant throughout $\tau=0.2$ to $\tau=0.7$. However, its estimated coefficient decreases from $2.75$ ($\tau=0.2$) to $1.39$ ($\tau=0.7$), indicating its heterogeneous impacts on survival, i.e., stronger protective effect at lower quantiles and vice versa. 
	
	The effects of some SNPs are nearly zero for higher quantiles. For example, the estimated coefficient of AX.15207405\_G decreases from $2.03$ ($\tau=0.2$; $p=10^{-24}$) to $-0.05$ ($\tau=0.7$; $p=0.92$), with the estimated standard error increasing from $0.20$ to $0.48$. Similarly, the estimated coefficient of AX.40182999\_A decreases from $1.5$ ($\tau=0.2$; $p=9.6\times 10^{-13}$) to $-0.01$ ($\tau=0.7$; $p=0.96$). The results again hint at heterogeneous SNP effects in various risk groups, which cannot be detected using traditional Cox models.

	Finally, our results shed light on the roles of SNPs {in} the high-risk group (i.e., lower quantiles). Specifically, we map the $83$ SNPs with significant effects at the $0.2$-th quantile by Fused-HDCQR  to the corresponding genes and rank the genes by the  number of significant SNPs (over total number of SNPs for each gene in the parenthesis), which are  TP53 (14/321), RRM1 (14/174), ERCC1 (10/167), BRCA1 (10/114), ALK (8/163), ROS1 (5/294), EGFR (5/261), ERBB2 (4/167),
	and 6 other genes with numbers of significant SNPs less than 4. While these genes were reported to be associated with lung cancer 
	\citep{toyooka2003tp53,takeuchi2012ret,rosell2011pretreatment,lord2002low,zheng2007dna,sasaki2006egfr,brose2002braf}, our analysis provides more detailed information as to which SNPs and locations of the genes are jointly associated with the lung cancer survival, as well as the estimated effects and  uncertainties. Analysis of heterogeneous SNP effects has been gaining increasing research attention in lung cancer research \citep{mckay2017large,dong2012potentially,huang2009genome}, and  beyond it \citep{garcia2008heterogeneity,cheng2010prostate,gulati2014systematic}.
	
	\section{Conclusions}\label{ch3.7}
	
	%Motivated by analyzing survival data with censoring and a large number of potential predictors, we have proposed a framework based on high dimensional censored quantile regressions for both estimation and inference of the model parameters. Censored quantile regression is a good alternative to Cox proportional hazards model, as it models various quantiles of the outcome distribution and is powerful in detecting heterogeneous effects at different quantiles.
	
	%The Fused-HDCQR procedure consists of two components, fused coefficient estimator and its variance estimator. The fused estimator is the average of estimators from multiple random sample splits and selections. The model-free variance estimator is derived using functional delta method and the re-sampling properties. The globally concerned CQR framework models a fine grid of quantiles that covers an interval of interest, which provides comprehensive understanding of the outcome distribution, as well as precise inferences for each predictor at different quantiles. As illustrated in Figure \ref{plotcoefs}, such statistical inferences tell much more information of the regression model comparing to simple variable selection. Furthermore, machine learning tasks like prediction and classification are byproducts of our method as a parsimonious model is available from our estimation and inference result. 
	
	Our proposed procedure involves  repeated estimates from low dimensional CQRs, which are computationally straightforward and can be {efficiently} implemented with parallel computing. We require the variable selection to possess a sure screening property as in condition \ref{a4}. This seems to be supported  by our simulations, which find our procedure works well when the variable selection method can  select a superset of the true model with high probability. Our condition is much weaker than a condition of selection consistency as specified in \cite{fei2019drawing}.
	
	For the selection of $B$, we recommend $B$ to be in the same order of the sample size $n$. Smaller $B$ might not affect coefficient estimation much; but it might yield inaccurate estimated standard errors, leading to incorrect inferences.  In addition, we opt to define $\Gamma_m$ by setting the grid as $n/\log p$ equally spaced points between $\tau_0$ and $\tau_U$. This may cover the quantile interval well, with reasonable computation efficiency. 
	
	There are  open  questions  to be addressed. First, substantial work is needed for handling highly correlated predictors as the performance of our method, like the other competing methods, deteriorates when  correlations among predictors become stronger.
	Second, it is  of  interest to investigate an alternative method when the sparsity condition fails. For example, it is challenging  to find an effective strategy to draw inference when a non-negligible portion of predictors have small but non-zero effects. We will pursue them elsewhere. %%, in which case grouping methods might be desired \citep{kang2017partition}
	
	%We conclude that our procedure could also be extended to other types of censoring, for example left truncation, as well as modeling time-dependent covariates or time-varying effects.

		\section*{Acknowledgements} We are deeply grateful toward the Editor, the AE
		and the two referees for their constructive comments and suggestions that have improved the manuscript substantially. 
		We thank our long-term collaborator, David Christiani of Harvard Medical School, for providing the BLCSC data. The work is partially supported by  grants from NIH (5R01CA249096 and U01CA209414).
		
	\bibliography{mybib}
	\bibliographystyle{chicago}

	\newpage
	\pagenumbering{gobble}

 %%%%%%%%

 \begin{table}[H]
  \begin{threeparttable}
\centering
\caption{Summary of  variable selection results based on the  simulated datasets.%	 \yl{Perhaps it would be best to keep the table as below as this table is about comparing the variable screening methods (not the fused method). You may want to do read the response to	 AE Q1(1). Please also copy paste this to replace Table 1 in the main text.}
%	 W12, \cite{wang2012quantile};
%	 F14, \cite{fan2014adaptive}. %Note that \cite{wang2012quantile,fan2014adaptive} are only implemented and compared in Example 1.
\label{simtab1}}
{\begin{tabular}{lrrrrrrrrr}
  \hline
  && & & \multicolumn{3}{c}{TP} & \multicolumn{3}{c}{FP}  \\
 &$(n,p)$& CR & $q$ & L-HDCQR  & M09 & F09 &  L-HDCQR  & M09 & F09 \\ 
  \hline
\multirow{2}{*}{Example 1}
&(300,1000) & 0.25 & 3 & 2.67 & 2.12 & 1.64 & 7.95 & 0.00 & 0.19  \\ 
&(700,1000) & 0.25 & 3 & 2.98 & 2.78 & 2.27 & 13.08 & 0.01 & 0.34  \\ 
\multirow{2}{*}{Example 2}
&(300,1000) & 0.22 & 4 & 3.60 & 3.58 & 2.22 & 12.45 & 0.00 & 0.22 \\ 
&(700,1000) & 0.23 & 4 & 3.99 & 3.99 & 3.54 & 11.29 & 0.00 & 0.64  \\ 
\multirow{3}{*}{Example 3}
&(300,1000) & 0.20 & 5 & 3.82 & 3.63 & 1.91 & 10.00 & 0.00 & 0.17 \\ 
&(700,1000) & 0.20 & 5 & 4.81 & 4.77 & 4.35 & 11.73 & 0.01 & 0.54 \\ 
&(700,2000) & 0.19 & 5 & 4.78 & 4.76 & 4.17 & 16.34 & 0.00 & 0.47 \\    \hline
\end{tabular}}
\begin{tablenotes}
      \small
      \item Note:  CR, average censoring rate;
      $q = |S^*|$;
      TP, average true positives;
      FP, average false positives;
	 M09, \cite{meinshausen2009p};
	 F09, \cite{fan2009ultrahigh};
	 L-HDCQR,  \cite{zheng2018high}.
    \end{tablenotes}
  \end{threeparttable}
\end{table}
%%%%%%%%%%%%%%%%%%%%%%%%%%%%%%%%%

\begin{table}[H]
  \begin{threeparttable}
%%\large
	\caption{Results of Example 1  based on the  simulated datasets. 
	\label{tab1}}
\centering
{\begin{tabular}{lrrrrrrrrrr}
  \hline
 & \multicolumn{5}{c}{Bias} & EmpSD & SE &  Cov & \multicolumn{2}{c}{Power} \\ \cline{2-6}\cline{10-11}
 & Oracle & Fused & Z18 & F14 & W12 &  \multicolumn{3}{c}{Fused} & Fused & M09 \\ 
  \hline
  & \multicolumn{10}{c}{$n=300,~p=1000$}\\
\multirow{3}{*}{$\beta_{21}=0.5$} & 0.02 & 0.02 & -0.38 & -0.50 & -0.48 & 0.14 & 0.13 & 0.93 & 0.97 & 0.06 \\ 
   & 0.02 & 0.01 & -0.24 & -0.49 & -0.48 & 0.12 &0.13  & 0.95 & 0.98 & 0.04 \\ 
   & 0.01 & 0.01 & -0.13 & -0.50 & -0.48 & 0.12 & 0.13& 0.96 & 1.00 & 0.02 \\ 
  \multirow{3}{*}{$\beta_{41}=1$}  & -0.01 & -0.01 & -0.02 & -0.91 & -0.33 & 0.14 & 0.13 & 0.92 & 1.00 & 0.99 \\ 
   & -0.00 & -0.00 & -0.03 & -0.68 & -0.32 & 0.14& 0.12 &  0.92 & 1.00 & 0.98 \\ 
   & 0.02 & 0.01 & -0.01 & -0.70 & -0.30 &0.17 & 0.14  & 0.93 & 1.00 & 0.94 \\ 
  \multirow{3}{*}{$\beta_{61}=1.5$} & -0.00 & 0.01 & 0.00 & -0.92 & -0.24 &0.12 &0.13  & 0.92 & 1.00 & 1.00 \\ 
  & 0.00 & 0.01 & 0.01 & -0.64 & -0.25 & 0.11 & 0.13 &  0.97 & 1.00 & 1.00 \\ 
   & 0.02 & 0.01 & 0.02 & -0.70 & -0.25 & 0.13 &0.14  & 0.95 & 1.00 & 1.00 \\ 
  &\multicolumn{10}{c}{$n=700,~p=1000$}\\
  \multirow{3}{*}{$\beta_{21}=0.5$} & -0.02 & -0.01 & -0.01 & -0.47 & -0.23 &0.09 & 0.08&  0.92 & 1.00 & 0.56 \\ 
  & -0.01 & -0.01 & -0.01 & -0.39 & -0.22 &0.08 & 0.08& 0.89 & 1.00 & 0.65 \\ 
  & -0.01 & -0.01 & -0.01 & -0.40 & -0.23 & 0.10& 0.09& 0.89 & 1.00 & 0.44 \\ 
  \multirow{3}{*}{$\beta_{41}=1$} & 0.00 & 0.00 & 0.04 & -0.53 & -0.17 &0.09 &0.08 & 0.91 & 1.00 & 1.00 \\ 
  & -0.00 & 0.00 & 0.03 & -0.49 & -0.19 & 0.09 &0.08 & 0.90 & 1.00 & 1.00 \\ 
  & -0.01 & -0.01 & 0.01 & -0.53 & -0.18 &0.08 & 0.10& 0.87 & 1.00 & 1.00 \\ 
 \multirow{3}{*}{$\beta_{61}=1.5$} & 0.01 & 0.01 & 0.06 & -0.54 & -0.21 & 0.10& 0.09& 0.93 & 1.00 & 1.00 \\ 
  & 0.01 & 0.01 & 0.03 & -0.62 & -0.21 & 0.08& 0.08 & 0.93 & 1.00 & 1.00 \\ 
  & -0.00 & 0.00 & 0.03 & -0.71 & -0.21 & 0.07 & 0.09 & 0.94 & 1.00 & 1.00 \\ 
%%  0's &  & -0.00 & 0.00 & -0.00 & 0.00 & 0.93 & 0.07 & 0.00 \\ 
%%  &  & -0.00 & 0.00 & 0.00 & 0.00 & 0.93 & 0.07 & 0.00 \\ 
%%  &  & -0.00 & 0.00 & 0.00 & 0.00 & 0.92 & 0.08 & 0.00 \\ 
   \hline
\end{tabular}}
\begin{tablenotes}
      \small
      \item Note: %% See the footnote of Table~\ref{simtab1}. 
      Each $\beta$ has three rows corresponding to   $\tau=.25,.5,.75$ from the top to bottom;
EmpSD, empirical standard deviation; SE, average standard error; Cov, coverage probability; Oracle, Oracle estimator;	 Z18,  \cite{zheng2018high}.F14,  \cite{fan2014adaptive}; W12, \cite{wang2012quantile}; M09, \cite{meinshausen2009p}.
    \end{tablenotes}
  \end{threeparttable}
\end{table}

\begin{table}[H]
  \begin{threeparttable}
\centering
\caption{Results of Example 2  based on the simulated datasets.\label{tab2}}
	\begin{tabular}{lrrrrrrrrr}
  \hline
 & \multicolumn{3}{c}{Bias} & EmpSD & SE & Cov & Freq & \multicolumn{2}{c}{Power} \\ \cline{2-4}\cline{9-10}
 & Oracle & Fused & Z18  & \multicolumn{4}{c}{Fused} & Fused & M09 \\ 
  \hline
  & \multicolumn{8}{c}{$n=300,~p=1000$}\\
  \multirow{3}{*}{$\beta_{4}= 1.5Q_\varepsilon(\tau)$}  & 0.01 & 0.13 & 0.29 &0.32& 0.31 & 0.88 &\multirow{3}{*}{0.73} & 0.82 & 0.16 \\ 
  & -0.05 & -0.07 & 0.06 &0.33 & 0.29 & 0.90 & & 0.11 & 0.00 \\ 
  & 0.01 & -0.14 & -0.05 & 0.31 & 0.34 & 0.82 & & 0.62 & 0.10 \\ 
  \multirow{3}{*}{$\beta_{21}=1$} & -0.01 & -0.01 & -0.01 &0.14 &0.13 & 0.90 &\multirow{3}{*}{0.69} & 1.00 & 0.88 \\ 
   & -0.03 & -0.01 & -0.05  &0.12 &0.12 & 0.91 & & 1.00 & 0.92 \\ 
   & -0.01 & -0.00 & -0.02 &0.14& 0.13 &0.92 & & 1.00 & 0.84 \\ 
  \multirow{3}{*}{$\beta_{41}=1.5$} & 0.01 & 0.01 & 0.03 &0.13 &0.13& 0.90 & \multirow{3}{*}{0.99}& 1.00 & 1.00 \\ 
   & -0.01 & 0.01 & 0.03 & 0.12 & 0.13 &0.93 & & 1.00 & 1.00 \\ 
   & -0.00 & 0.02 & -0.02 & 0.13 & 0.14 &0.93 & & 1.00 & 1.00 \\ 
  \multirow{3}{*}{$\beta_{61}=2$} & -0.03 & -0.03 & 0.04 &0.13 &0.13 & 0.91 & \multirow{3}{*}{1.00}& 1.00 & 1.00 \\ 
  & -0.03 & -0.02 & 0.03 &  0.11 & 0.13 &0.92 & & 1.00 & 1.00 \\ 
   & -0.01 & -0.01 & -0.00 & 0.12 & 0.15 & 0.95 & & 1.00 & 1.00 \\ 
 %% 0's &  & 0.00 & 0.00 & 0.93 & 0.07 & 0.00 \\ 
  %% &  & 0.00 & 0.00 & 0.93 & 0.07 & 0.00 \\ 
  %% &  & 0.00 & 0.00 & 0.93 & 0.07 & 0.00 \\ 
  \hline
  & \multicolumn{8}{c}{$n=700,~p=1000$}\\
  \multirow{3}{*}{$\beta_{4} = 1.5Q_\varepsilon(\tau)$}  & 0.03 & 0.08 & 0.19 & 0.19 & 0.21 & 0.92 & \multirow{3}{*}{0.89} & 0.99 & 0.61 \\ 
   & 0.02 & 0.03 & 0.14 & 0.18 & 0.19 & 0.89 & & 0.11 & 0.00 \\ 
  & 0.04 & -0.03 & -0.01 & 0.21 & 0.23 &0.92 & & 0.97 & 0.56 \\ 
  \multirow{3}{*}{$\beta_{21}=1$} & 0.01 & 0.01 & 0.05 & 0.09 & 0.08 &0.94 &\multirow{3}{*}{0.99} & 1.00 & 1.00 \\ 
   & 0.01 & 0.01 & 0.01 & 0.08 & 0.08 &0.87 & & 1.00 & 1.00 \\ 
  & 0.01 & 0.01 & 0.05 & 0.10 &0.09 &0.89 & & 1.00 & 1.00 \\ 
  \multirow{3}{*}{$\beta_{41}=1.5$} & -0.01 & 0.00 & 0.08 & 0.08&  0.08 & 0.94 & \multirow{3}{*}{1.00}& 1.00 & 1.00 \\ 
   & -0.00 & 0.00 & 0.05 & 0.09 & 0.08& 0.92 & & 1.00 & 1.00 \\ 
   & 0.00 & 0.01 & 0.04 & 0.09 & 0.09 &0.95 & & 1.00 & 1.00 \\ 
  \multirow{3}{*}{$\beta_{61}=2$} & -0.01 & -0.01 & 0.10 & 0.08 & 0.09 & 0.93 &\multirow{3}{*}{1.00} & 1.00 & 1.00 \\ 
   & -0.01 & -0.01 & 0.06 & 0.08 & 0.09 &0.91 & & 1.00 & 1.00 \\ 
   & -0.00 & -0.00 & 0.07 & 0.09 & 0.10 & 0.90 & & 1.00 & 1.00 \\ 
 %% 0's &  & -0.00 & 0.00 & 0.93 & 0.07 & 0.00 \\ 
 %%  &  & -0.00 & 0.00 & 0.93 & 0.07 & 0.00 \\ 
 %% &  & -0.00 & 0.00 & 0.93 & 0.07 & 0.00 \\ 
   \hline
\end{tabular}
\begin{tablenotes}
      \small
      \item Note: See the  footnote of Table~\ref{tab1}; 
 Freq, average selection frequency in $B$ splits.
    \end{tablenotes}
  \end{threeparttable}
\end{table}

\begin{table}[H]
  \begin{threeparttable}
\centering
\small
\caption{Results of Example 3   based on the simulated datasets. \label{tab3}}
\begin{tabular}{lrrrrrrrrr}
  \hline
 & \multicolumn{3}{c}{Bias} & EmpSD & SE &  Cov & Freq & \multicolumn{2}{c}{Power} \\\cline{2-4}\cline{9-10}
 & Oracle & Fused & Z18 & \multicolumn{4}{c}{Fused} & Fused & M09 \\ 
  \hline
 & \multicolumn{8}{c}{$n=300,p=1000$}\\
\multirow{3}{*}{$\beta_{2} = \phi_1(\tau)$}  & 0.08 & 0.06 & 0.59 &0.34 & 0.36 & 0.94 &\multirow{3}{*}{0.71} & 0.06 & 0.00 \\ 
  & 0.34 & 0.37 & 1.01 & 0.52 & 0.51 &0.89 & & 0.20 & 0.00 \\ 
  & 0.08 & -0.20 & -0.05 & 0.80 & 0.72 &0.89 & & 0.87 & 0.06 \\ 
  \multirow{3}{*}{$\beta_{5} = \phi_{4}(\tau)$}  & 0.10 & 0.14 & 0.27 & 0.65 & 0.50 & 0.90 &\multirow{3}{*}{0.67} & 0.77 & 0.36 \\ 
  & -0.16 & -0.20 & -0.36 &0.62 & 0.51 & 0.91 & & 0.19 & 0.00 \\ 
  & 0.02 & 0.06 & -0.03 &0.56 & 0.52 & 0.90 & & 0.10 & 0.00 \\ 
  {$\beta_{21}=1.5$} & 0.02 & 0.03 & 0.04 & 0.25 & 0.23 &0.95 &0.65 & 1.00 & 0.77 \\ 
  {$\beta_{41}=2$} & 0.01 & -0.00 & 0.02 & 0.23 & 0.25 &0.93 & 0.93& 1.00 & 0.99 \\ 
  {$\beta_{61}=2.5$} & -0.02 & 0.07 & 0.19 &0.21 & 0.26 & 0.94 & 0.99& 1.00 & 1.00 \\ 
 %% 0's &  & 0.00 & 0.00 & 0.93 & 0.07 & 0.00 \\ 
  \hline
  & \multicolumn{8}{c}{$n=700,p=1000$}\\
  \multirow{3}{*}{$\beta_{2} = \phi_1(\tau)$} & 0.01 & 0.04 & 0.27 &0.21 & 0.23& 0.94 &\multirow{3}{*}{0.96} & 0.06 & 0.00 \\ 
  & 0.13 & 0.30 & 0.79 & 0.37 & 0.40 & 0.88 & & 0.27 & 0.01 \\ 
  & 0.00 & 0.08 & 0.35 & 0.51 & 0.51 & 0.90 & & 1.00 & 0.77 \\ 
  \multirow{3}{*}{$\beta_{5} = \phi_{4}(\tau)$}  & 0.06 & 0.09 & 0.18 &0.33 & 0.33& 0.91 &\multirow{3}{*}{0.92} & 0.99 & 0.92 \\ 
  & -0.09 & -0.19 & -0.23 & 0.35 & 0.34 & 0.85 & & 0.21 & 0.00 \\ 
  & -0.01 & -0.04 & -0.08 & 0.37 & 0.31 & 0.94 & & 0.06 & 0.00 \\ 
  {$\beta_{21}=1.5$} & -0.00 & 0.00 & 0.04 &0.16 & 0.17 & 0.97 & 0.98 & 1.00 & 1.00 \\ 
  {$\beta_{41}=2$} & 0.01 & -0.02 & -0.01 & 0.15 & 0.18 &0.95 & 1.00 & 1.00 & 1.00 \\ 
  {$\beta_{61}=2.5$} & 0.01 & 0.00 & 0.07 & 0.18 & 0.18 &0.94 & 1.00 & 1.00 & 1.00 \\ 
  %%0's &  & -0.00 & 0.00 & 0.93 & 0.07 & 0.00 \\ 
  \hline
  & \multicolumn{8}{c}{$n=700,p=2000$}\\
  \multirow{3}{*}{$\beta_{2} = \phi_{1}(\tau)$} & -0.01 & 0.05 & 0.13 &0.32 &0.32& 0.93 &\multirow{3}{*}{0.93} & 0.07 & 0.00 \\ 
  & 0.10 & 0.26 & 0.59 & 0.46 &0.44 &0.91 & & 0.09 & 0.02 \\ 
  & 0.05 & -0.07 & 0.15 & 0.53 & 0.46 & 0.87 & & 0.74 & 0.58 \\ 
  \multirow{3}{*}{$\beta_{5} = \phi_{4}(\tau)$} & 0.10 & 0.10 & 0.25 &0.45 & 0.35  &0.84 &\multirow{3}{*}{0.90} & 1.00 & 0.83 \\ 
   & -0.03 & -0.18 & -0.31 &0.41 & 0.36 & 0.89 & & 0.76 & 0.01 \\ 
   & -0.00 & -0.01 & -0.13 &0.36& 0.34 & 0.85 & & 0.15 & 0.00 \\ 
  {$\beta_{21}=1.5$} & 0.01 & 0.01 & 0.03 & 0.18 & 0.21 &0.98 &0.98 & 1.00 & 1.00 \\ 
  {$\beta_{41}=2$} & 0.01 & 0.02 & -0.07 &0.22 & 0.20 & 0.91 & 0.99& 1.00 & 0.98 \\ 
  {$\beta_{61}=2.5$} & -0.01 & -0.01 & -0.05 & 0.25 & 0.20 &0.94 & 1.00& 1.00 & 0.98 \\ 
  %%0's &  & 0.00 & 0.00 & 0.94 & 0.06 & 0.00 \\ 
   \hline
\end{tabular}
\begin{tablenotes}
      \small
      \item Note: See the  footnote of Tables~\ref{tab1} and \ref{tab2};  For  $\beta_{2}$ and $\beta_{5}$,  the numbers are shown at $\tau=.25,.5,.75$  from the top to the bottom and, for the other $\beta$'s, at $\tau = 0.5$.
          \end{tablenotes}
  \end{threeparttable}
\end{table}

\begin{table}[H]
  \begin{threeparttable}
\caption{Comparisons of  computing time  (on average per dataset in seconds) when performing Example 1.
	 \label{simtab11}}
\begin{tabular}{rrrrrr}
  \hline
 & Fused & Z18 & W12 & F14 & M09 \\ 
  \hline
$(n,p)=(300,1000)$ & 888 & 853 & 509 & 390 & 170 \\ 
$(n,p)=(700,1000)$ & 3,108 & 1,812 & 2,230 & 1,231 & 440 \\ 
 % Example 2\\
%$(300,1000)$ & 898 & 899 & - & - & 170 \\ 
%$(700,1000)$ & 3200 & 1925 & - & - & 447 \\ 
 % Example 3\\
%$(300,1000)$ & 892 & 960 & - & - & 208 \\ 
%$(700,1000)$ & 3258 & 1950 &-  & - & 496 \\ 
%$(700,2000)$ & 9803 & 3310 & - & - %& 816 \\ 
   \hline
\end{tabular}
\begin{tablenotes}
      \small
      \item Note: see the  footnote  of Table~\ref{tab1}.
    \end{tablenotes}
  \end{threeparttable}
\end{table}

	\begin{table}
		\centering
		\caption{Patients' characteristics in the BLCSC samples. \\ ($n=674$)\label{dem_blcs}}
		\begin{tabular}{llr}
			\hline
			   &&  Mean (SD) \\
			Age& & 60 (10.8) \\	
			&& Count (\%)  \\
			Female && 259 (38)  \\
			Education level	&$\le$ High school  & 264 (39)\\
			&$>$ High school & 410 (61) \\
			Smoking & Non-active & 418 (62)  \\
			&Active & 256 (38)  \\
			Cancer type & 		Adenocarcinoma & 283 (42) \\
			&	Squamous cell  & 110 (16)  \\
			&	Other  & 281 (42)   \\
			Cancer stage  & 1  & 283 (42) \\
			&  2  & 110 (16)\\
			& 3  & 256 (38)\\
			&4  & 25 (4)\\
			\hline
		\end{tabular}
	\end{table}

	\begin{sidewaystable}[ht]
		\centering
		\caption{Analysis of the BLCSC  data with Fused-HDCQR. The SNPs are sorted by their $p$-values at $\tau=0.2$, corresponding to the high-risk groups. Results for  the top 10 and the bottom 3 are presented. \label{tab_blcs}}
		{\begin{tabular}{rrrrgggrrr}
				\hline
				& Estimator & SE & $p$-value & Estimator & SE & $p$-value & Estimator & SE & $p$-value \\ 
				$\tau$ & \multicolumn{3}{c}{0.2} & \multicolumn{3}{c}{0.3}& \multicolumn{3}{c}{0.4} \\
				\hline
				Int & 6.90 & 0.25 & 1.4E-165 & 7.48 & 0.28 & 4.3E-157 & 7.94 & 0.24 & 3.2E-241 \\ 
				Adeno & 0.20 & 0.16 & 2.1E-01 & 0.14 & 0.18 & 4.5E-01 & 0.02 & 0.13 & 8.7E-01 \\ 
				Squamous & -0.16 & 0.16 & 3.0E-01 & -0.20 & 0.16 & 2.1E-01 & -0.34 & 0.13 & 1.0E-02 \\ 
				Stage2 & -0.82 & 0.24 & 6.3E-04 & -0.99 & 0.25 & 6.0E-05 & -0.98 & 0.24 & 3.2E-05 \\ 
				Stage3 & -0.97 & 0.17 & 1.6E-08 & -1.04 & 0.20 & 2.0E-07 & -1.13 & 0.14 & 2.0E-15 \\ 
				Stage4 & -1.54 & 0.17 & 3.0E-20 & -1.77 & 0.20 & 1.7E-19 & -1.86 & 0.14 & 2.2E-42 \\ 
				Age & -0.01 & 0.01 & 1.5E-02 & -0.01 & 0.01 & 3.0E-02 & -0.02 & 0.01 & 1.0E-02 \\ 
				Edu & 0.08 & 0.14 & 6.0E-01 & 0.06 & 0.15 & 6.9E-01 & 0.07 & 0.13 & 5.8E-01 \\ 
				Female & -0.30 & 0.13 & 2.2E-02 & -0.35 & 0.14 & 1.0E-02 & -0.37 & 0.12 & 1.6E-03 \\ 
				Smoke & -0.42 & 0.13 & 1.1E-03 & -0.48 & 0.14 & 5.0E-04 & -0.52 & 0.11 & 3.4E-06 \\ 
				AX.37793583\_T & 2.75 & 0.22 & 3.0E-36 & 2.61 & 0.20 & 4.6E-39 & 2.39 & 0.20 & 3.7E-33 \\ 
				AX.83104700\_A & 2.32 & 0.20 & 4.0E-31 & 1.91 & 0.19 & 6.3E-24 & 1.54 & 0.19 & 1.5E-15 \\ 
				AX.15207405\_G & 2.03 & 0.20 & 1.0E-24 & 1.59 & 0.22 & 9.8E-13 & 1.17 & 0.21 & 3.7E-08 \\ 
				AX.16619495\_T & 1.79 & 0.20 & 3.3E-19 & 1.36 & 0.20 & 1.3E-11 & 0.97 & 0.20 & 1.2E-06 \\ 
				AX.13920550\_G & 1.93 & 0.23 & 2.5E-17 & 1.41 & 0.28 & 5.3E-07 & 0.87 & 0.27 & 1.6E-03 \\ 
				AX.83444620\_C & 1.39 & 0.17 & 7.4E-17 & 1.05 & 0.19 & 6.6E-08 & 0.71 & 0.21 & 8.8E-04 \\ 
				AX.82902859\_T & 1.58 & 0.20 & 8.7E-16 & 1.19 & 0.18 & 2.0E-11 & 0.90 & 0.12 & 3.4E-14 \\ 
				AX.40182999\_A & 1.50 & 0.21 & 9.6E-13 & 1.01 & 0.25 & 3.9E-05 & 0.64 & 0.14 & 6.5E-06 \\ 
				AX.82976133\_A & 2.32 & 0.33 & 3.8E-12 & 2.02 & 0.35 & 6.7E-09 & 1.58 & 0.35 & 6.1E-06 \\ 
				AX.82900605\_G & 2.21 & 0.35 & 1.6E-10 & 1.91 & 0.29 & 9.1E-11 & 1.54 & 0.33 & 2.9E-06 \\ 
				... &&&&&&&&& \\
				AX.41828883\_G & 1.4E-03 & 0.34 & 1.00 & -3.2E-02 & 0.42 & 0.94 & -5.7E-02 & 0.54 & 0.92 \\ 
				AX.11293250\_T & -3.6E-04 & 0.14 & 1.00 & 6.2E-02 & 0.15 & 0.67 & 5.0E-02 & 0.12 & 0.68 \\ 
				AX.37863475\_C & -3.1E-04 & 0.26 & 1.00 & -1.1E-01 & 0.25 & 0.68 & -1.8E-01 & 0.24 & 0.46 \\ 
				\hline
		\end{tabular}}
	\end{sidewaystable}
	\begin{sidewaystable}[ht]
		\centering
		\caption*{Analysis of the BLCSC data with Fused-HDCQR, continued.}
		{	\begin{tabular}{rrrrgggrrr}
				\hline
				& Estimator& SE & $p$-value & Estimator& SE & $p$-value & Estimator& SE & $p$-value \\ 
				$\tau$ & \multicolumn{3}{c}{0.5} & \multicolumn{3}{c}{0.6}& \multicolumn{3}{c}{0.7} \\
				\hline
				Int & 8.30 & 0.27 & 4.8E-214 & 8.55 & 0.30 & 4.9E-180 & 8.69 & 0.35 & 2.8E-132 \\ 
				Adeno & -0.09 & 0.15 & 5.3E-01 & -0.09 & 0.13 & 4.8E-01 & -0.09 & 0.13 & 5.1E-01 \\ 
				Squamous & -0.50 & 0.15 & 1.0E-03 & -0.60 & 0.16 & 2.1E-04 & -0.50 & 0.19 & 7.1E-03 \\ 
				Stage2 & -0.88 & 0.25 & 5.0E-04 & -0.73 & 0.24 & 2.1E-03 & -0.57 & 0.19 & 2.8E-03 \\ 
				Stage3 & -1.08 & 0.17 & 1.7E-10 & -0.91 & 0.15 & 6.4E-10 & -0.68 & 0.16 & 2.0E-05 \\ 
				Stage4 & -1.91 & 0.15 & 7.0E-38 & -1.93 & 0.14 & 1.7E-44 & -1.69 & 0.16 & 2.1E-27 \\ 
				Age & -0.02 & 0.00 & 3.3E-05 & -0.02 & 0.01 & 1.3E-03 & -0.02 & 0.01 & 1.9E-03 \\ 
				Edu & 0.15 & 0.14 & 2.7E-01 & 0.16 & 0.13 & 2.2E-01 & 0.11 & 0.13 & 4.0E-01 \\ 
				Female & -0.44 & 0.11 & 6.4E-05 & -0.47 & 0.12 & 1.6E-04 & -0.38 & 0.15 & 1.3E-02 \\ 
				Smoke & -0.53 & 0.15 & 4.9E-04 & -0.36 & 0.16 & 2.4E-02 & -0.31 & 0.15 & 3.8E-02 \\ 
				AX.37793583\_T & 2.16 & 0.20 & 4.1E-28 & 1.84 & 0.28 & 2.8E-11 & 1.39 & 0.25 & 4.2E-08 \\ 
				AX.83104700\_A & 1.15 & 0.27 & 1.6E-05 & 0.58 & 0.27 & 3.5E-02 & 0.13 & 0.25 & 6.0E-01 \\ 
				AX.15207405\_G & 0.75 & 0.25 & 2.3E-03 & 0.34 & 0.37 & 3.5E-01 & -0.05 & 0.48 & 9.2E-01 \\ 
				AX.16619495\_T & 0.66 & 0.22 & 3.1E-03 & 0.44 & 0.31 & 1.5E-01 & 0.18 & 0.35 & 6.1E-01 \\ 
				AX.13920550\_G & 0.54 & 0.27 & 4.3E-02 & 0.26 & 0.60 & 6.7E-01 & 0.11 & 0.60 & 8.6E-01 \\ 
				AX.83444620\_C & 0.55 & 0.23 & 2.0E-02 & 0.29 & 0.22 & 1.8E-01 & 0.01 & 0.18 & 9.7E-01 \\ 
				AX.82902859\_T & 0.73 & 0.13 & 4.2E-08 & 0.51 & 0.32 & 1.1E-01 & 0.22 & 0.46 & 6.3E-01 \\ 
				AX.40182999\_A & 0.41 & 0.18 & 2.6E-02 & 0.22 & 0.27 & 4.1E-01 & -0.01 & 0.30 & 9.6E-01 \\ 
				AX.82976133\_A & 1.17 & 0.42 & 5.4E-03 & 0.61 & 0.52 & 2.4E-01 & 0.24 & 0.46 & 6.0E-01 \\ 
				AX.82900605\_G & 1.22 & 0.35 & 4.5E-04 & 0.86 & 0.34 & 1.1E-02 & 0.50 & 0.31 & 1.0E-01 \\ 
				...&&&&&&&&&\\
				AX.41828883\_G & 0.26 & 0.60 & 0.66 & 0.32 & 0.52 & 0.54 & 0.12 & 0.68 & 0.86 \\ 
				AX.11293250\_T & -0.00 & 0.12 & 1.00 & -0.09 & 0.12 & 0.44 & -0.09 & 0.15 & 0.56 \\ 
				AX.37863475\_C & -0.24 & 0.20 & 0.23 & -0.37 & 0.17 & 0.03 & -0.57 & 0.32 & 0.08 \\ 
				\hline
		\end{tabular}}
	\end{sidewaystable}

\end{document}